\documentclass[useAMS,usenatbib]{mn2e}
\usepackage{psfig, epsf, epsfig}
\title[Planetary nebula systems in galaxies]{Stellar halos
and elliptical galaxy formation: Origin of dynamical properties of
the planetary nebular systems}
\author[Kenji Bekki and Eric Peng]{Kenji Bekki$^{1}$\thanks{E-mail:
bekki@bat.phys.unsw.edu.au} 
and Eric W. Peng$^{2}$\\
$^{1}$School of Physics, University of New South Wales,
              Sydney 2052, NSW, Australia\\
$^{2}$Herzberg Institute of Astrophysics, 5071 West Saanich Road,
	Victoria, BC V8V 2X6, Canada}
\begin{document}

\date{Accepted, Received 2005 February 20; in original form }

\pagerange{\pageref{firstpage}--\pageref{lastpage}} \pubyear{2005}

\maketitle

\label{firstpage}

\begin{abstract}

Recent spectroscopic observations of planetary nebulae (PNe)
in several elliptical galaxies have revealed structural and kinematical
properties of the outer stellar halo regions.
In order to elucidate the origin of the properties of these planetary
nebula systems (PNSs),  
we consider the merger scenario  in which an elliptical galaxy
is formed by merging of  spiral galaxies.
Using numerical simulations,
we particularly investigate radial profiles of projected PNe number densities,
rotational velocities, and velocity dispersions of PNSs 
extending to the outer halo regions 
of elliptical galaxies formed from major and unequal-mass merging.
We find that the radial profiles of the project number densities
can be fitted to the power-law
and the mean number density in the outer halos of the ellipticals
can be  more than 
an order of magnitude higher than that of the original spiral's halo.
The PNSs are found to 
show a significant amount of rotation ($V/\sigma$ $>$ $0.5$)
in the outer halo regions ($R$ $>$ $5R_{\rm e}$) of the ellipticals.
Two-dimensional velocity fields of PNSs are derived from the simulations
and their dependences on model parameters of galaxy merging are discussed
in detail. 
We compare the simulated kinematics of PNSs with that of  
the PNS observed in NGC 5128 and thereby discuss advantages
and disadvantages of the merger model in explaining the observed
kinematics  of the PNS.  
We also find that the kinematics of PNSs in elliptical galaxies
are quite diverse depending on the orbital configurations of galaxy merging,
the mass ratio of merger progenitor spirals, 
and the viewing angle of the galaxies.  This variation translates
directly into possible biases by a factor of two in observational
mass estimation.  However, the biases in the total mass estimates can
be even larger.  The best case systems viewed edge-on can appear to
have masses lower than their true mass by a factor of 5,
which suggests that
current observational studies  on PNe kinematics  of elliptical galaxies
can significantly underestimate their real masses.

\end{abstract}


\section{Introduction}

Structural, kinematical, and chemical properties of stars in the outer
regions of galaxies, their halos, provide vital clues to their global
formation and evolutionary histories.
Although the physical properties of stellar halos
have been extensively investigated 
mostly for spiral (e.g., the Galaxy and the M31)
and dwarf galaxies in the Local Group
(e.g., Eggen, Lynden-Bell, \& Sandage 1962;
Mould \& Kristian 1986; Norris 1986; Freeman 1987;
Durrell, Harris, \& Pritchet 1994, 2001;
Pritchet \& van den Bergh 1998; 
Reitzel, Guhathakurta, \& Gould 1998; Grillmair et al. 1996;
Chiba \& Yoshii 1998; Davidge 2002),
recent observations have started to reveal physical properties of
stellar halos in giant elliptical galaxies, all of which are beyond the
Local Group 
(e.g., Soria et al. 1996; Harris, Harris, \& Poole 1999 hereafter referred to as HHP;
Harris \& Harris 2000, HH00; Harris \& Harris 2002, HH02; Marleau et al. 2000;
Rejkuba et al. 2002; Gregg et al. 2004).
These data prompt us to investigate how the properties of stellar halos
can be used to understand the formation of giant elliptical galaxies.

Beyond projected radii of  $\sim$ $2R_{\rm e}$, it becomes very
difficult to study the integrated properties of stars in ellipticals
because of the low stellar surface brightness.  Hence, studies of
stellar halos tend to rely on resolving individual stars or star clusters. 
While there are significant observational difficulties in revealing the
three dimensional structure and kinematics of halo red
giants and AGB stars, it has been possible to use color-magnitude
diagrams to derive
their metallicity distribution function (hereafter MDF)
and constrain the past star formation 
and chemical evolution histories of their host elliptical galaxies
(e.g., HHP; HH00, HH02). 
Kinematic properties of these stars, however,
which can contain vital clues to both structure and formation of ellipticals, 
are still extremely difficult to obtain.
Furthermore, these previous observations have difficulties in deriving
{\it  global} stellar halo  distributions, 
which are also considered to have valuable information on
the past merging histories of galaxies (e.g., fossil tidal streams),  
mainly because the halo fields investigated cover
only a minor portion of the entire halo.

Planetary nebulae (PNe) are a valuable complement to integrated light
and red giant studies.  They are powerful tools
for studying the dynamical states of elliptical galaxies,
because we can more readily identify PNe through their bright
[O III] emissions lines and typically measure their radial velocities
to an accuracy of $\sim$ 15 km s$^{-1}$ (e.g., Peng et al. 2004a, PFF04a).
Spectroscopic observations of the radial velocities of PNe in 
several nearby elliptical galaxies therefore have succeeded in
deriving global mass distributions and velocity fields of these galaxies
(Ciardullo et al. 1993; Hui et al. 1995; Arnaboldi et al. 1998;
Mendez et al. 2001; Romanowsky et al. 2003; Napolitano et al. 2004;
PFF04a).
For example, kinematic information of PNe in the stellar halo of NGC 5128
at radii up to $\sim$ 80 kpc allows the authors to 
derive the zero velocity curve (ZVC) in the two-dimensional velocity
field of the PNS
and thereby to discuss the triaxial shape of the mass distribution
of this galaxy (PFF04a).
This observed diversity in kinematics of PNSs
is providing fresh clues to the origin of early-type galaxies.
(Romanowsky et al. 2003; PFF04a).

In spite of this importance of PNe studies in elliptical galaxies,
only a few theoretical and numerical attempts have been made to discuss
dynamical properties of PNSs in elliptical galaxies.
For example, using kinematical data up to $\sim$ $6R_{\rm e}$ for
an heterogeneous sample of elliptical galaxies,
Napolitano et al. (2004) derived the dependences
of the radial gradients of the mass-to-light-ratios
on the $B-$band magnitude of the galaxies 
and thereby discussed whether the derived dependences
can be consistent with elliptical galaxy formation models based on
a $\Lambda$CDM model.
However, PNSs provide a very rich data set from which to study
ellipticals.  In particular, we are interested in exploring 
{\it the 2D distributions of structural and kinematical properties
of PNSs}, which can provide some vital clues both to the triaxial
shapes of the global mass distributions of galaxies 
and to their formation processes. 
The origin of 2D structural and kinematical properties of PNSs 
extending to the outer halo regions of 
elliptical galaxies remains unclear.
Furthermore providing theoretical  predictions on dynamical properties of PNSs
will help to interpret the properties of PNSs that will be obtained 
in future systematic observations for elliptical galaxies.

The purpose of this paper is thus to investigate structural and kinematical
properties of PNSs of elliptical galaxies
based on numerical simulations.
In order to elucidate the origin of the observed properties of the PNSs, 
we adopt the merger scenario (Toomre 1977) in which
elliptical galaxies are proposed to be formed by major merging
of two spiral galaxies.
The present numerical investigation is two-fold as follows.
We first describes radial profiles of structural and kinematical properties
of PNSs and two-dimensional (2D) velocity fields of PNSs in elliptical galaxies
and their dependences on model parameters.
We then compare the simulated kinematics of the PNSs  with the corresponding
observations for NGC 5128 and thereby try to provide the best model for 
the PNS in NGC 5128 and discuss advantages and disadvantages of the model
in explaining the kinematics of the PNS self-consistently.
The essential reason for our choice of the NGC 5128 PNS is that
PFF04a have recently investigated radial velocities
of the NGC 5128's 780 PNe (among 1141 PNe), 
which represents the largest kinematical study
of an elliptical galaxy to date and thus can be compared with our simulations
in the most self-consistent manner. 
Furthermore, the present study is complimentary to
those by  Bekki et al. (2003, BHH03) and Beasley et al. (2003)
which numerically and semi-analytically investigated 
the MDF of the stellar halo of NGC 5128 
but did not investigate the dynamical properties.

In the present paper,
we mainly investigate structural and kinematic properties of
PNSs for a large radial extent in elliptical galaxies
(0 $\le$ $R$ $\le$ $10R_{\rm e}$), which include the outer faint
stellar halos and the main bodies of the galaxies.
Although previous numerical simulations investigated 
spatial distributions and the line-of-sight velocity
distributions of stars in elliptical and S0 galaxies formed from merging  for
$R$ $<$  $2.5R_{\rm e}$ 
(e.g., Bendo \& Barnes 2000; Cretten et al. 2001; Naab \& Trujillo 2005),
they did not investigate the dynamical properties for the outer halo regions
of elliptical galaxies. Therefore, the present numerical results 
on the dynamical properties of the outer stellar halos including PNe 
($5R_{\rm e}$  $\le$ $R$ $\le$ $10R_{\rm e}$) may provide   
new clues to elliptical galaxy formation.
Physical properties of these halo stars may well have fossil information
on angular momentum redistribution of stars  
(or conversion from galactic orbital angular momentum into intrinsic
angular momentum)
which is an  essential physical process of galaxy merging.
We therefore can expect that dynamical properties of stellar halos,
which can be provided by studies of PNSs, may depend strongly on physics of
galaxy merging.  Through the course of this paper, we thus interpret our
simulations in the context of the capabilities of present-day PN surveys
in order to facilitate the comparison between theory and
observations.

The plan of the paper is as follows: In the next section,
we describe our numerical models  for the formation of PNSs extending to
the outer halo regions 
in galaxy mergers. 
In \S 3, we 
present the numerical results
mainly on the final 2D distributions  of structural and kinematical properties
in  merger remnants (i.e., elliptical galaxies) 
for variously different merger models.
In \S 4, we compare the simulated results of PNSs with
observations of the PNS in NGC 5128 (Cen A). 
In \S 5, we discuss whether physical properties of stellar halos
can give any constraints on galaxy formation, based on the 
present numerical results.
In this section, we also discuss possible different 
properties of PNSs in galaxies with different Hubble types. 
We summarize our  conclusions in \S 6.

\begin{table*}
\centering
\caption{Model parameters}
\begin{tabular}{ccccccccccl}
Model no. & 
Orbital type & 
$m_{2}$   & 
Comments \\
M1 & -- & -- & isolated disk  \\
M2 & HI & 1.0 & fiducial  model\\
M3 & PP & 1.0 &   \\
M4 & RR & 1.0 &   \\
M5 & LA & 1.0 &   \\
M6 & BO & 1.0 &  \\
M7 & HI & 0.1 &   \\
M8 & HI & 0.3 &   \\
M9 & PO & 0.5 & NGC 5128 model \\
M10 & -- & 1.0 & Multiple merger  \\
\end{tabular}
\end{table*}

\section{Model}

\subsection{Merger models}

We investigate the dynamical evolution of 
fully self-gravitating galaxies 
composed of stars and dark matter via collisionless numerical simulations carried out 
on a GRAPE board (Sugimoto et al. 1990).  
Since our numerical methods for modeling dynamical evolution 
of  galaxy mergers have already been described by Bekki \& Shioya
(1998) and by BHH03,  we give only a brief
review here.  
The total mass and the size of a disk of the merger progenitor spiral
are $M_{\rm d}$ and $R_{\rm d}$, respectively. 
Henceforth, all masses and lengths are measured in units
of $M_{\rm d}$ and $R_{\rm d}$, respectively, unless
specified. Velocity and time are measured in units of $v$ = $
(GM_{\rm d}/R_{\rm d})^{1/2}$ and $t_{\rm dyn}$ = $(R_{\rm
d}^{3}/GM_{\rm d})^{1/2}$, respectively, where $G$ is the
gravitational constant and assumed to be 1.0 in the present
study. If we adopt $M_{\rm d}$ = 6.0 $\times$ $10^{10}$ $ \rm
M_{\odot}$ and $R_{\rm d}$ = 17.5 kpc as a fiducial value, then
$v$ = 1.21 $\times$ $10^{2}$ km/s and $t_{\rm dyn}$ = 1.41
$\times$ $10^{8}$ yr, respectively. 
The merger progenitor spiral is composed of a dark matter halo,
a stellar disk, a stellar bulge, and a stellar halo. 

The mass ratio of the dark matter halo to the stellar disk
in a disk model is
fixed at 10 for all models.
We adopt the density distribution of the NFW 
halo (Navarro, Frenk \& White 1996) suggested from CDM simulations:
\begin{equation}
{\rho}(r)=\frac{\rho_{0}}{(r/r_{\rm s})(1+r/r_{\rm s})^2},
\end{equation} 
where  $r$, $\rho_{0}$, and $r_{\rm s}$ are
the spherical radius,  the central density of a dark halo,  and the scale
length of the halo, respectively.  
The value of $r_{\rm s}$ (0.6 in our units for $c$ = 10) is chosen such that
the rotation curve of a disk is reasonably consistent with observations.
The mass fraction  and
the scale length of the stellar bulge  
are fixed at 0.17 
(i.e., 17 \% of the stellar disk) 
and 0.04 (i.e., 20 \%) of the scale length of the stellar disk),
respectively, which are consistent with those of the bulge
model of the Galaxy.
The initial total mass of the stellar halo  is 0.01 in our units
for all models and has a number density 
distribution ($\rho (r) \sim r^{-3.5}$) like that of the stellar halo 
of the Milky Way (e.g., Chiba \& Beers 2000).

The total  number of collisionless particles  in each simulation is 112680 for
pair mergers and 281700 for multiple merger (described later).
The total particle number for dark matter halo, stellar disk, bulge,
stellar halo, and old halo GCs are 30000, 20000, 3340, 2000, and 1000,
respectively, for a disk.
Gravitational softening lengths in the GRAPE simulations are set to be
fixed at 0.15 (in our units) 
for the dark matter, 0.02 for the stellar disk and bulge,
0.09 for the stellar halo for each simulation.
Since we mainly analyze structural and kinematical properties of PNSs
for the scale of $1-100$kpc in galaxies, 
the resolution of the simulations ($\sim 350$ pc) is quite reasonable.

We investigate both pair mergers with the number 
of the progenitor spirals ($N_{\rm g}$) equal to 2
and multiple ones with $N_{\rm g}$ = 5. 
Since our main interest here is the formation of elliptical galaxies by pair mergers,
the details of the models and the results for the multiple mergers
are given in the Appendix B. 
In the
simulation of a pair merger model, the orbit of the
two spirals is set to be initially in the $x$-$y$ plane and the
distance between the center of mass of the two spirals ($r_{\rm
in}$) is 6$R_{\rm d}$ (105 kpc).
The pericentre distance ($r_{\rm p}$) and the orbital
eccentricity ($e_{\rm p}$) are assumed to be free parameters
which control orbital angular momentum and energy of the merging
galaxies.  For most merger models, $r_{\rm p}$ and
$e_{\rm p}$ are set to be 1.0 (in our units) and 1.0,
respectively. 
The spin of each galaxy in a merging and
galaxy is specified by two angles $\theta_{i}$ and
$\phi_{i}$ (in units of degrees), where the suffix $i$ is used to
identify each galaxy.  Here, $\theta_{i}$ is the angle between
the $z$-axis and the vector of the angular momentum of the disk,
and $\phi_{i}$ is the azimuthal angle measured from $x$ axis to
the projection of the angular momentum vector of the disk onto
the $x$-$y$ plane. 

We specifically show the following five
different and representative
models with different disk inclinations with respect to the
orbital plane: A prograde-prograde model represented by ``PP''
with $\theta_{1}$ = 0, $\theta_{2}$ = 30, $\phi_{1}$ = 0, and
$\phi_{2}$ = 0, 
a retrograde-retrograde (``RR'') with $\theta_{1}$ = 180,
$\theta_{2}$ = 150, $\phi_{1}$ = 0, and $\phi_{2}$ = 0, and a highly
inclined model (``HI'') with $\theta_{1}$ = 30, $\theta_{2}$ =
120, $\phi_{1}$ = 90, and $\phi_{2}$ = 180. 
The model with lower orbital angular
momentum ($r_{\rm p}$ = 0.2) and a HI orbital configuration is
labeled as ``LA'' whereas the model with bound orbit with $e_{\rm p}$ of 0.7
and a HI orbital configuration  is labeled as  ``BO''. 

  The total masses of elliptical galaxies can be underestimated, if
kinematical properties of PNSs are used for the mass estimation 
(e.g., Dekel et al. 2005). 
We also find that the masses of ellipticals can be significantly underestimated
if radial velocity profiles of PNSs are used under the assumption of 
the virial theorem. 
We just briefly discuss this point in the discussion section \S 5.3, because our main
focus is on 2D density and velocity fields of PNSs in elliptical galaxies.

\begin{figure}
\psfig{file=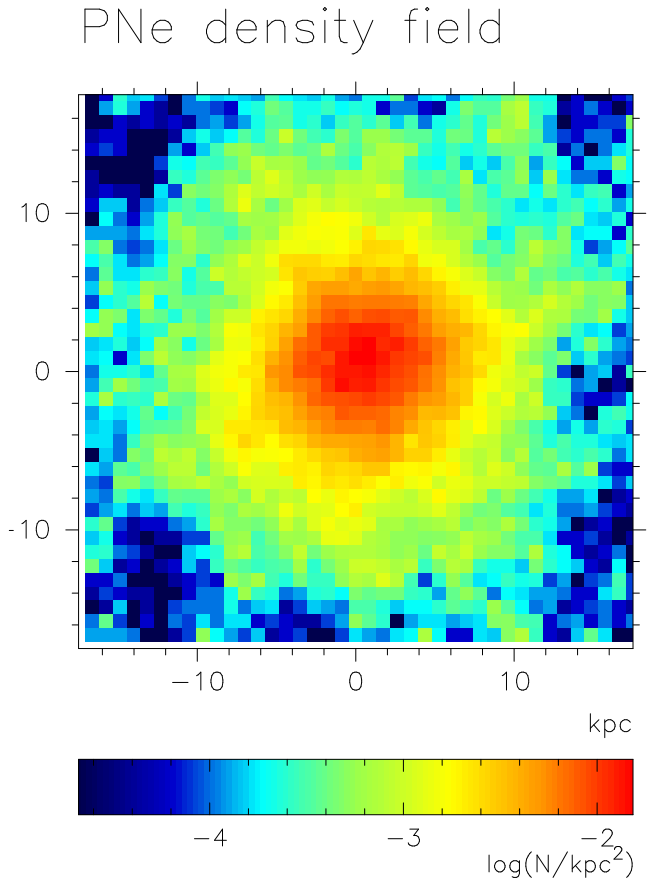,width=8.5cm}
\caption{ 
The two-dimensional (2D)  PNe density field of the stellar halo
in the isolated disk model M1 projected onto the $X$-$Y$ plane.
The stellar halo region with the size of 17.5 kpc is divided into
40 $\times$ 40 cells and
the projected PNe density at each cell is estimated by assuming
${\alpha}_{\rm PNe}$ of 9.4 $\times 10^{-9}$  PNe per $L_{\odot}$ in $B$-band
(Ciardullo et al. 1989).
The original spherical distribution of the PNS is clearly
seen in this smoothed 2D density field.
}
\label{Figure. 1}
\end{figure}

\begin{figure}
\psfig{file=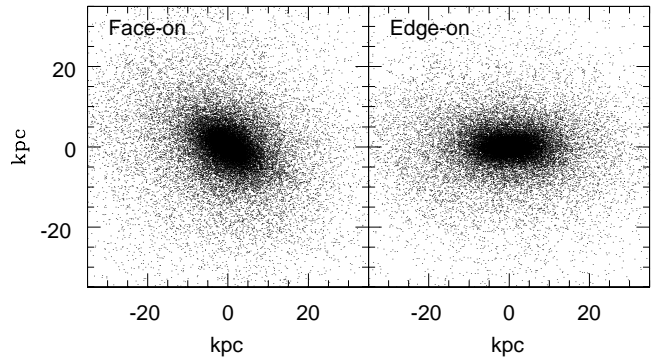,width=8.5cm}
\caption{ 
 The mass distributions of stars projected onto the $X$-$Y$ plane
(face-on view, left) and the $X$-$Z$ one (edge-on view, right)
for the fiducial model M2 at $T$ = 4.5 Gyr.
}
\label{Figure. 2}
\end{figure}

\subsection{Main points of analysis}

We mainly investigate 2D density and velocity fields of PNSs and compare
the results with the corresponding observations in a fully self-consistent
manner. We accordingly use Gaussian smoothing similar to that adopted in
observations (e.g., PFF04a) in order to derive smoothed density and velocity
fields of the simulated PNSs. The details of the methods to derive the
smoothed fields are given in the Appendix A.  
Figure 1 shows how the smoothed PNe density field looks like for the stellar  
halo consisting only of 1000 discrete stars in an isolated disk model M1
(viewed from face-on).

Although the MDFs of stellar halos (thus PNe) can give some constraints
on any theory of elliptical galaxies formation (e.g., HH00; HH02; 
BHH03), we do not intend to discuss them in this paper.
This is firstly because  (1) the importance of the MDFs have been already
discussed in previous papers (e.g., BHH03) 
and secondly because there is no currently available data set of the MDFs
of PNe for nearby giant elliptical galaxies. 
Instead, we mainly investigate (1) radial profiles of projected PNe number
densities, (2) 2D fields of line-of-sight velocity ($V_{\rm los}$) and
velocity dispersion ($\sigma$),  (3) rotation curve profiles ($V_{\rm rot}$),
and (4) radial profiles of $\sigma$ and $V_{\rm rot}/\sigma$,
for the PNSs of merger remnants.
Although we investigated the above four points for 32 merger models
with different $m_{2}$ and  orbital configurations, we show the 
results of 9 representative models among them.
Table 2 summarizes  the model parameters used for these 9 models. 
We show the results of the models at $T$ = 4.5 Gyr (at final time step),
where the time $T$ represents time that has elapsed since
the merger progenitor disks begin to merge.

We firstly show the results of the ``fiducial model'', model M2,
which show typical behaviors of stellar halo (thus PNS) formation 
in major galaxy merging (\S 3.1). The results of this model can be 
regarded as generic ones for elliptical galaxies formed from major merging.
Secondly we show the parameter dependences of the results in \S 3.2
with special emphasis on the dependence of projected PNe number 
densities and the 2D velocity fields on orbital configuration
and $m_{2}$. Thirdly, we present the best model which
can reproduce the most reasonably well
a number of dynamical properties of the PNS observed in
NGC 5128 (\S 4.1). It should be stressed here that 
the best model is the best model {\it among the present 32 models},
thus is not the one which reproduce all of the observed properties
of the PNS fully self-consistently. 
We try to understand advantages and disadvantages of the present
best model by comparing it with observations, 
and discuss whether (and how) physical effects that are not included
in the present models (e.g., star formation) 
can  be important for more successful reproduction
of the observations. 
The best model is model M9 with $e_{\rm p}$ = 1.0,
$r_{\rm p}$ = 0.05 in our units, 
$\theta_{1}$ = 0, $\theta_{2}$ = 80, $\phi_{1}$ = 0, and
$\phi_{2}$ = 0, and $m_{2}$ = 0.5 (The orbital configuration
is hereafter referred to as ``PO'').

\begin{figure}
\psfig{file=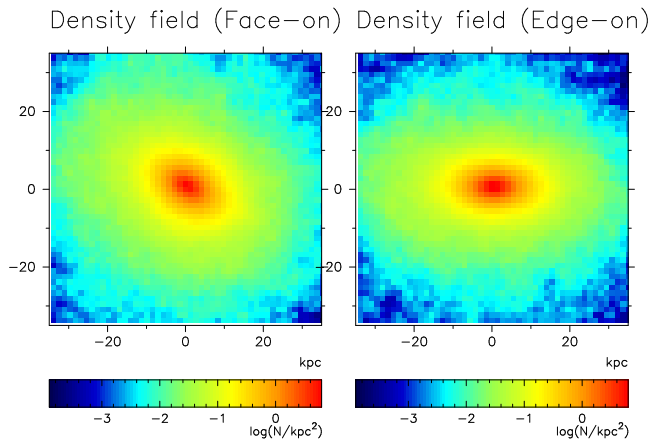,width=8.5cm}
\caption{ 
The 2D  PNe density field
for the face-on view
(left) and the edge-on view (right)
in the fiducial model at $T$ = 4.5 Gyr.
The 2D fields are produced based on the stellar mass distributions
shown in Figure 2.
}
\label{Figure. 3}
\end{figure}

\begin{figure}
\psfig{file=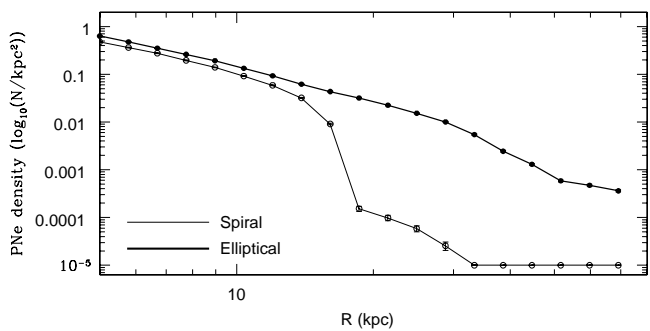,width=8.5cm}
\caption{ 
Radial profiles of the projected PNe number densities (${\rho}_{\rm PNe}$)
for the fiducial merger model M2 (thick)
and the isolated disk model M1 (thin).
Note that the elliptical galaxy formed in M2 shows significantly higher
${\rho}_{\rm PNe}$ in its outer region ($R$ $>$ 20 kpc) compared with
the spiral in M1.
}
\label{Figure. 4}
\end{figure}

\begin{figure}
\psfig{file=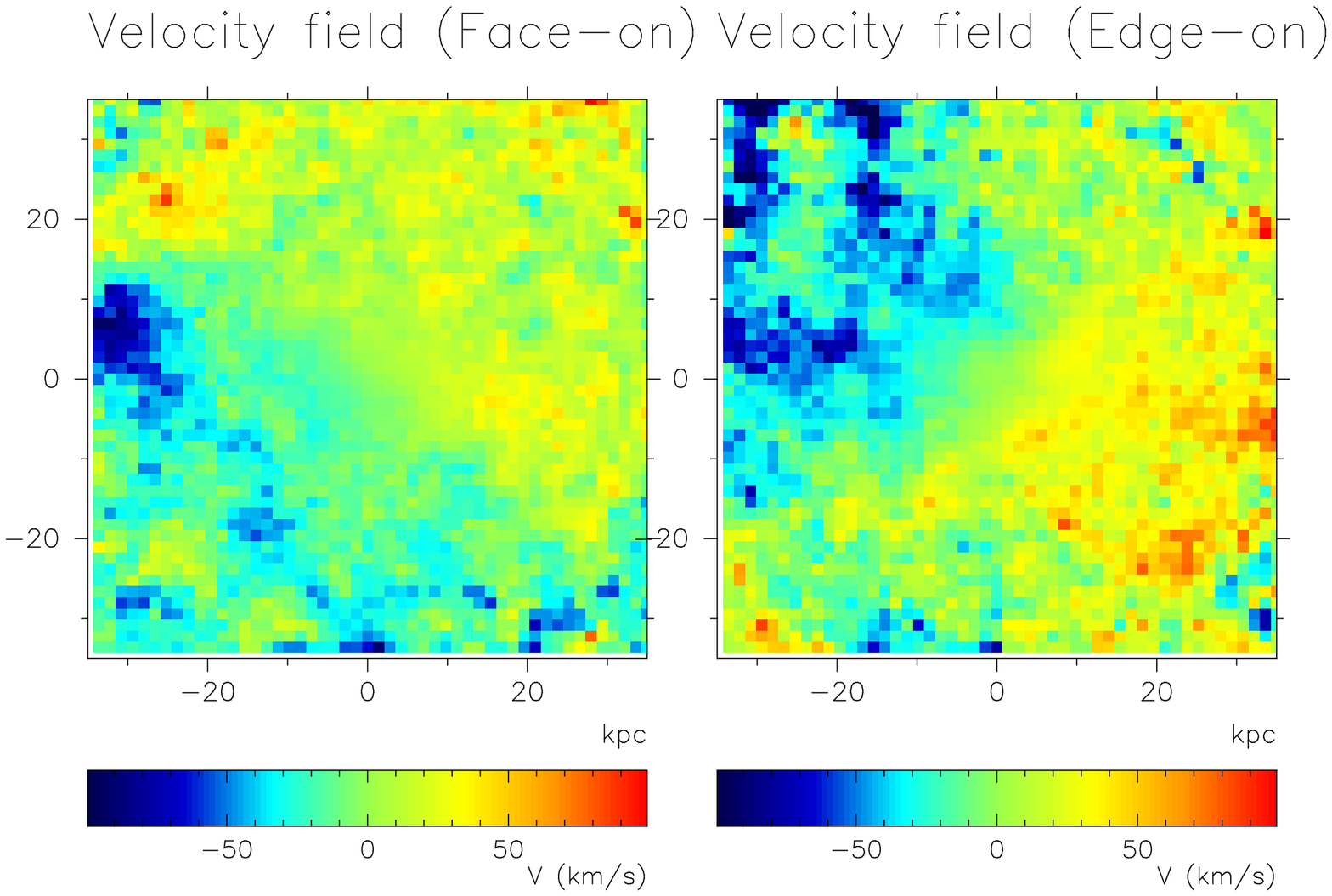,width=8.5cm}
\caption{ 
The 2D velocity fields for the face-on view (left)
and for the edge-one one (right)
for the fiducial model M2 at $T$ = 4.5 Gyr.
}
\label{Figure. 5}
\end{figure}

\begin{figure}
\psfig{file=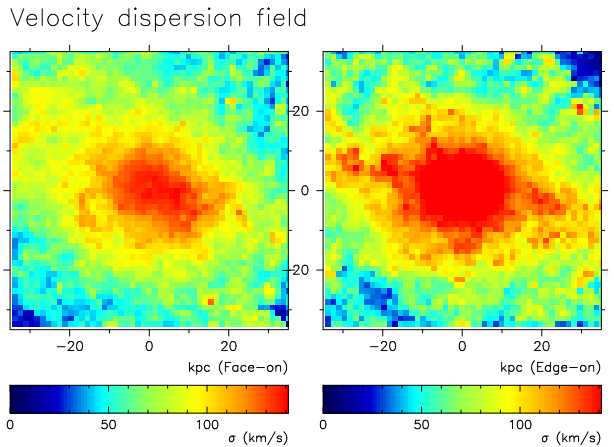,width=8.5cm}
\caption{ 
The same as Figure 5 but for the 2D velocity dispersion fields.
}
\label{Figure. 6}
\end{figure}

\section{Results}

\subsection{The fiducial model}

\subsubsection{Density fields}

Figure 2 shows the projected  stellar mass distributions of the fiducial
model M2 at $T$ = 4.5 Gyr 
in which the final merger remnant  shows  $R_{\rm e}$  of 5.2 kpc,
$L_{\rm B}$ = 3.7 $\times$ $10^{10}$ $L_{\odot}$, and
$N_{\rm PNe}$ = 350 within $\sim$ $5R_{\rm e}$.
The merger remnant  does not
show any clear signs of stellar  substructures 
(i.e.tidal tails and plums) in its halo region for
the two projections owing to the apparently completed dynamical relaxation 
by this time ($T$ = 4.5 Gyr).
It is also possible that the much less remarkable substructures are due to
the adopted smaller particle number (an order of $\sim 10^5$). 
The outer  stellar halo ($R$ $>$ $5R_{\rm e}$) in the edge-on view is more spherical 
($\epsilon$ $\sim$ 0.3) than the main body of the elliptical ($\epsilon$ $\sim$ 0.2 at
$R$ $\sim$ $2R_{\rm e}$).
The outer stellar halo is slightly more flattened than the dark matter halo
with $\epsilon$  = 0.15 at $R$ = $5R_{\rm e}$ in the edge-on view
and the major-axis of the stellar halo  nearly coincides with that of 
the dark matter halo.
The isophotal shape of this remnant formed from purely collisionless major galaxy
merging can show the negative sign of the $a_4$ parameter (e.g., Bekki \& Shioya 1997), 
and thus this remnant can be morphologically identified as a boxy elliptical galaxy.

Figures 3 and 4 describe the 2D density fields of the PNS in the elliptical 
and the radial profile of the PNe distribution, respectively.
Owing to the adopted assumption of constant ${\alpha}_{\rm PNe}$
(i.e., constant luminosity-specific number densities of PNe), 
the 2D density fields of the PNS  follows the stellar density ones 
of the elliptical and therefore shows a more flattened shape in the edge-on view.
The radial density profile is best fitted to the power-law with the exponent
of $-2.7$ for 0 $\le$ $R$ $\le$ 40 kpc ($\sim$ $8R_{\rm e}$),
which is slightly steeper than the observed slope of $-2.5$ for the PNS of NGC 5128
(PFF04a). The projected  PNe density of the elliptical
in its outer stellar halo ($R$ $>$ $5R_{\rm e}$) is 
more than two orders magnitudes higher than those of the merger progenitor spirals,
though the PNe densities in the central regions of galaxies are not
so different between the elliptical and the spirals.
The formation of such a high density stellar halo is demonstrated to be
closely associated with redistribution of disk stars through
tidal stripping of the stars during major galaxy merging (BHH).

Figure 5 shows the 2D fields of $V_{\rm los}$ (i.e., line-of-sight-velocity)
of the merger remnant for 0 $\le$ $R$ $\le$ 35 kpc ($\sim$ $7R_{\rm e}$) 
in the face-on and edge-on views. Clearly, the PNS shows a significant amount
of rotation ($\sim$ 100 km s$^{-1}$) along its major axis 
(i.e., the $X$-axis) for the edge-on view.  
Given the fact that the stellar halos of
the  merger progenitor spirals are assumed to have no net rotation,
the rotation in the outer halo of the elliptical can be considered to be obtained
during angular momentum redistribution of major galaxy merging.
This angular momentum transfer processes have been already well discussed
in previous papers (e.g., Barnes 1998 for a review).

The axis of rotation in the edge-on view
however does  not coincide with the minor axis of the PNe density
field shown in Figure 3: Radial velocity gradient can become maximum
if we measure it along a line connecting two points ($X$, $Z$) = ($-35$, 35)
and (35, $-35$) kpc for the edge-on 2D velocity field. 
Minor axis rotation can be clearly seen in the outer halo region of the elliptical
for the edge-on 2D velocity field and the velocity variation along the minor
axis (i.e., the $Z$-axis) is complicated. 
Also it should be stressed that the edge-on 2D field has  several regions with
large absolute magnitudes of $V_{\rm los}$ (e.g., ($X$, $Z$) = ($35$, $-8$) and
($-20$, $3$) kpc).
Such inhomogeneity in the velocity space  may indicate that 
dynamical relaxation is not  completely ended in a real term even a few Gyr after the coalescence 
of two spirals in the fiducial model: 
Fossil evidences of past violent relaxation can be proved in velocity structures of
outer stellar halos in elliptical galaxies.

The face-on 2D velocity field also shows rotation along the major and the minor
axes of the projected density distribution 
of the PNS, though the rotation in the outer stellar
halo for the face-on 2D field is less remarkable compared with the edge-on one.
The global appearance of the face-on 2D velocity field is smoother than
that of the edge-on one and complicated velocity variation
along the minor axis of the PNS can not be clearly seen. 
Thus these results imply  that (1) PNSs in elliptical galaxies
can show a significant amount of rotation in their outer halo regions,
(2) minor-axis rotation of PNSs is one of important characteristics of elliptical
galaxies,
and (3) the 2D velocity fields can be different for different viewing angles. 

\begin{figure}
\psfig{file=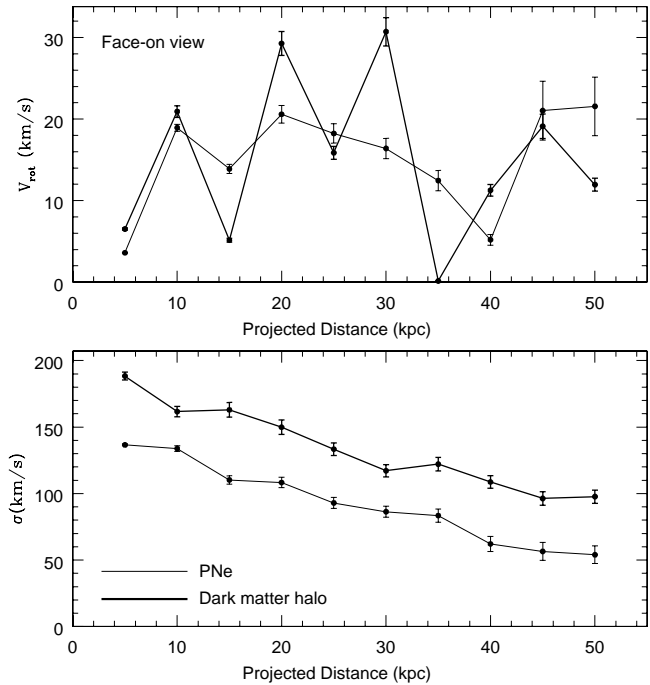,width=8.5cm}
\caption{ 
The radial profile of the rotation curve $V_{\rm rot}$ (upper) and the velocity
dispersion $\sigma$ (lower) for the dark matter halo (thick) and
for stars (thin) in the fiducial model M2 at $T$ = 4.5 Gyr.
$V_{\rm rot}$ and $\sigma$ are estimated for the model
projected onto the $X$-$Y$ plane (i.e.,  the face-on view).
The methods  to derive $V_{\rm rot}$ and $\sigma$ and their error bars
are described in detail in  the Appendix A.
}
\label{Figure. 7}
\end{figure}

\begin{figure}
\psfig{file=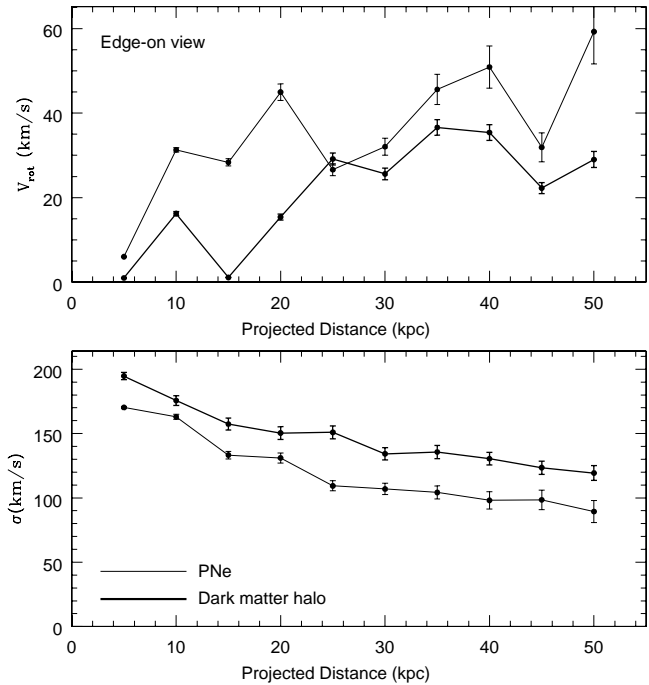,width=8.5cm}
\caption{ 
The same as Figure 7 but for the edge-on view (i.e., the $X$-$Z$ plane).
}
\label{Figure. 8}
\end{figure}

As shown in Figure 6, both the face-on and the edge-on 
2D velocity dispersion ($\sigma$) fields
show (1) higher velocity  dispersion in the inner regions of the elliptical,
(2) a shallower gradient of velocity dispersion along the major axis 
(i.e., the $X$-axis) compared with that along the minor one,
and (3) globally flattened shapes, in particular, in outer stellar halo regions.
These three can be regarded as important kinematical characteristics of
stellar halos (PNSs) in elliptical galaxies formed by major merging.
The flattened velocity dispersion fields are due to anisotropic velocity
dispersion in the triaxial mass distribution of the elliptical galaxy in this fiducial model.
The dispersion field appears to be more homogeneous in the face-on view
than in the edge-on one, though some remarkable 
substructures that can result from yet incomplete dynamical relaxation of this system 
can be seen in both projections.
The results shown in Figures 4,  5, and 6 clearly suggest that
even if the 2D density fields of PNSs appear to be quite regular and axisymmetric,
their 2D velocity and velocity dispersion fields can be more inhomogeneous
and less axisymmetric:
The kinematical properties of a stellar halo in an elliptical galaxy 
can provide some evidences of the past dynamical processes of its formation,
if penetrative analysis of such ``fossil record'' properties  can be done.

Figures 7 and 8 show the radial profiles of rotational velocities ($V_{\rm rot}$)
and velocity dispersions ($\sigma$) for the dark matter halo and the stellar
components (i.e., PNe) in the elliptical of the fiducial model. 
It is clear from these figures that (1) the dark matter halo shows a larger
velocity dispersion (by a factor of $1.1-2.0$, depending on radius) than
the PNe, (2) the PNe shows a significant amount of rotation $40-60$ km s$^{-1}$ 
in the entire halo region ($R$ $>$ $5R_{\rm e}$)
of the elliptical for the edge-on view,
(3) the maximum $V_{\rm rot}$ in the halo region is higher in the edge-on
view than in the face-on one,
(4) the radial gradient of $V_{\rm rot}$ is quite
different between different projections,
and (5) the radial velocity dispersion profiles 
(e.g., the central $\sigma$ and the gradient) 
do not differ significantly with each other.
The above result (1) strongly suggests that we can significantly underestimate
the total mass of an elliptical galaxy, 
if we use only the velocity dispersion data of the PNS and thereby 
derive the total mass based on the scalar virial theorem
in which the mass is linearly proportional to 
the product of $R_{\rm e}$ and $\sigma$:
We need to derive the total mass of an elliptical 
by  using both dispersion and rotation data of the PNS.
The result (2) is consistent with previous numerical simulations
(e.g., Heyl et al. 1996).

\begin{figure}
\psfig{file=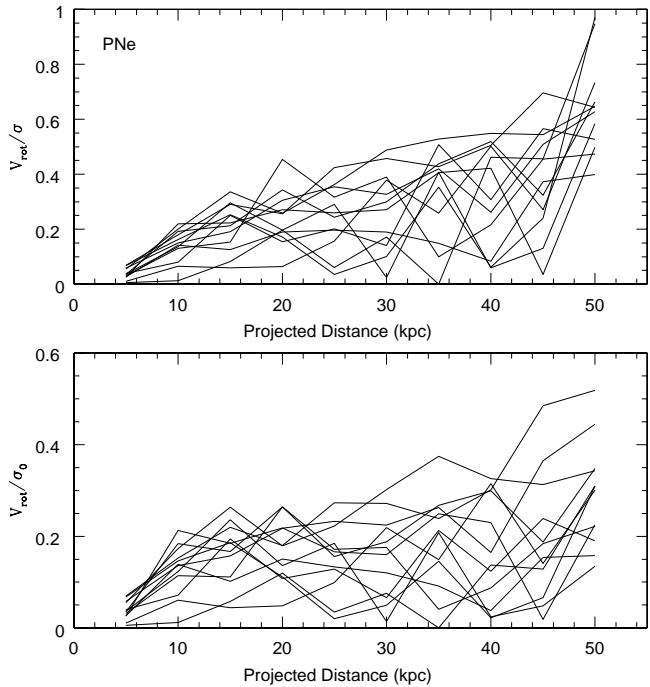,width=8.5cm}
\caption{ 
The radial profiles of $V_{\rm rot}/\sigma$ (upper)
and the normalized rotation curve
$V_{\rm rot}/{\sigma}_{0}$ (lower) for the fiducial
model viewed from  12 representative, and different viewing angles.
Note that there is a  trend of increasing $V_{\rm rot}/\sigma$
and $V_{\rm rot}/{\sigma}_{0}$ with radius.
}
\label{Figure. 9}
\end{figure}

\begin{figure}
\psfig{file=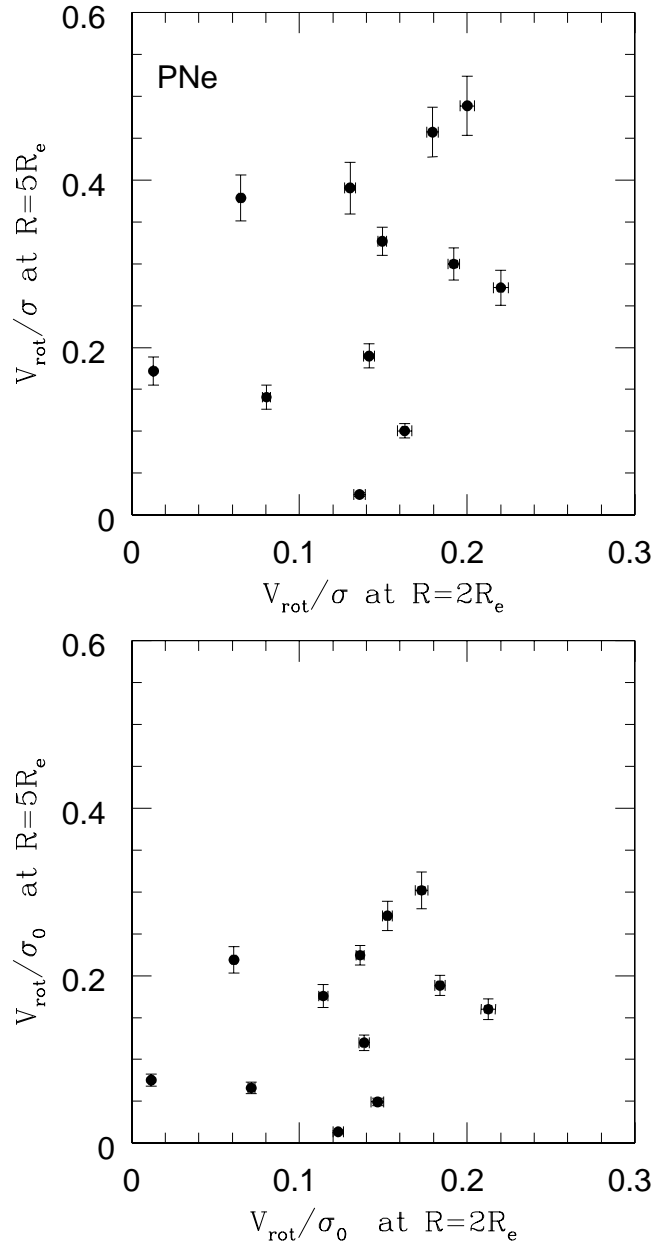,width=8.5cm}
\caption{ 
$V_{\rm rot}/\sigma$ at $R$ = $5R_{\rm e}$ as a function of
$V_{\rm rot}/\sigma$ at $R$ = $2R_{\rm e}$ (upper)
and $V_{\rm rot}/{\sigma}_{0}$  at $R$ = $5R_{\rm e}$
as a function of $V_{\rm rot}/{\sigma}_{0}$  at $R$ = $2R_{\rm e}$
(lower)
for the fiducial model viewed from 12 representative, and different directions.
}
\label{Figure. 10}
\end{figure}

Given the fact that the radial dispersion profile of the PNS decreases with radius
in the fiducial model,
the above result (2) means that the PNS shows a higher value of $V_{\rm rot}/\sigma$ 
in the outer halo of the elliptical.
(e.g.,  $V_{\rm rot}/\sigma$ of $\sim$ 0.7 at $R$ $\simeq$ $5R_{\rm e}$). 
This larger $V_{\rm rot}/\sigma$ also results from the redistribution of
angular momentum of disks stars during major merging 
(i.e.,   conversion of initial orbital angular
momentum of merger progenitor disks into final intrinsic one of the merger remnant),
as explained for Figure 5. 
This result suggests  that 
larger $V_{\rm rot}/\sigma$ in the outer stellar halo of an elliptical galaxy
can be a fossil evidence that the elliptical was formed from major galaxy merging. 
The derived PNS with larger $V_{\rm rot}/\sigma$ in the  outer halo region 
is in a striking contrast to the Galactic stellar halo with little rotation
(e.g., Freeman 1987), which implies that stellar halo kinematics can be 
significantly different between spirals and ellipticals.
Recent numerical simulations based on the currently favored 
cold dark matter (CDM) theory of galaxy formation 
have shown that the Galactic stellar halo with little rotation is formed 
at high redshift ($z$ $>$ 1) by
dissipative and dissipationless merging of smaller subgalactic clumps
and their resultant tidal disruption in the course 
of gravitational contraction of the Galaxy (Bekki \& Chiba 2000, 2001).
Thus the above result (2) implies that differences in
formation processes of stellar halos between different types of
galaxies (e.g., spirals vs ellipticals) can be reflected on
the stellar halo kinematics.
The above results (3) - (5) imply that radial profiles of rotational velocities
of PNSs in elliptical galaxies depend more strongly on viewing angles than
those of velocity dispersion.

Figure 9 describes how $V_{\rm rot}/\sigma$ and $V_{\rm rot}/{\sigma}_{0}$
depend on radius in the PNS of the fiducial model for 12 representative projections. 
There can be seen a trend of increasing $V_{\rm rot}/\sigma$ and $V_{\rm rot}/{\sigma}_{0}$
with increasing radius, though the radial profiles rise and fall 
with radius significantly.
The higher $V_{\rm rot}/\sigma$ and $V_{\rm rot}/{\sigma}_{0}$ of the PNS in the outer 
halo region of the elliptical for most of the projections (viewing angles)
means that the PNS in the outer halo region
is the most likely to be observed as being strongly supported by rotation. 
Figure 10 shows that a PNS with higher $V_{\rm rot}/\sigma$ at $R$ = $2R_{\rm e}$ 
is more likely to have  higher $V_{\rm rot}/\sigma$ at $R$ = $5R_{\rm e}$:
There is a correlation in stellar kinematics between the main body and the stellar halo
of the elliptical, though the correlation is weak.
Figure 10 also shows a stronger correlation 
between $V_{\rm rot}/{\sigma}_{0}$ at $R$ = $2R_{\rm e}$ 
and  $V_{\rm rot}/{\sigma}_{0}$ at $R$ = $5R_{\rm e}$.
The derived correlations imply that 
if elliptical galaxies are formed from major galaxy merging,
the kinematic of their main bodies
derived from integrated absorption spectra (for $R$ $<$ $2R_{\rm e}$) 
can be correlated with that of their outer stellar halos derived from 
radial velocity fields of PNe (for $R$ $>$ $5R_{\rm e}$).
The predicted ``halo-host kinematic correlation'' can be tested by
future extensive systematic studies of PNe kinematics for elliptical
galaxies with already known kinematics of the main bodies.

\begin{figure}
\psfig{file=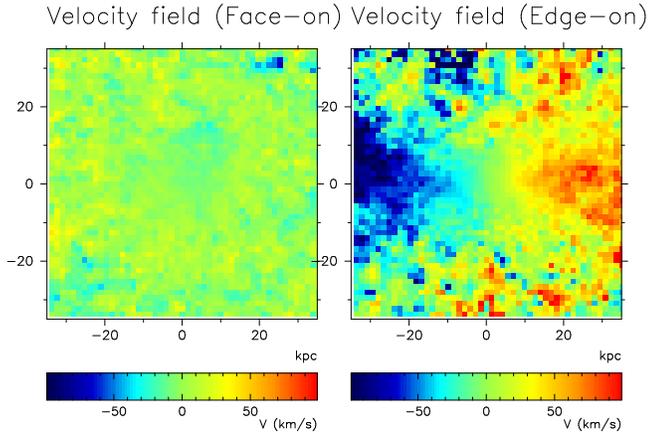,width=8.5cm}
\caption{ 
The 2D velocity fields for the face-on view (left)
and for the edge-one one (right)
for the model M3 at $T$ = 4.5 Gyr.
}
\label{Figure. 11}
\end{figure}

\begin{figure}
\psfig{file=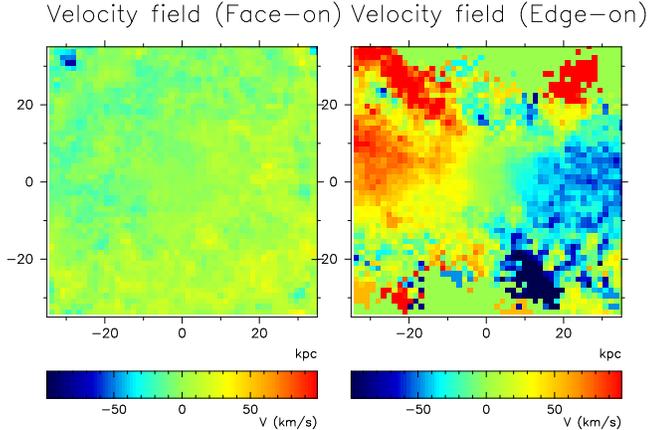,width=8.5cm}
\caption{ 
The same as Figure 11 but for the model M4.
}
\label{Figure. 12}
\end{figure}

\begin{figure}
\psfig{file=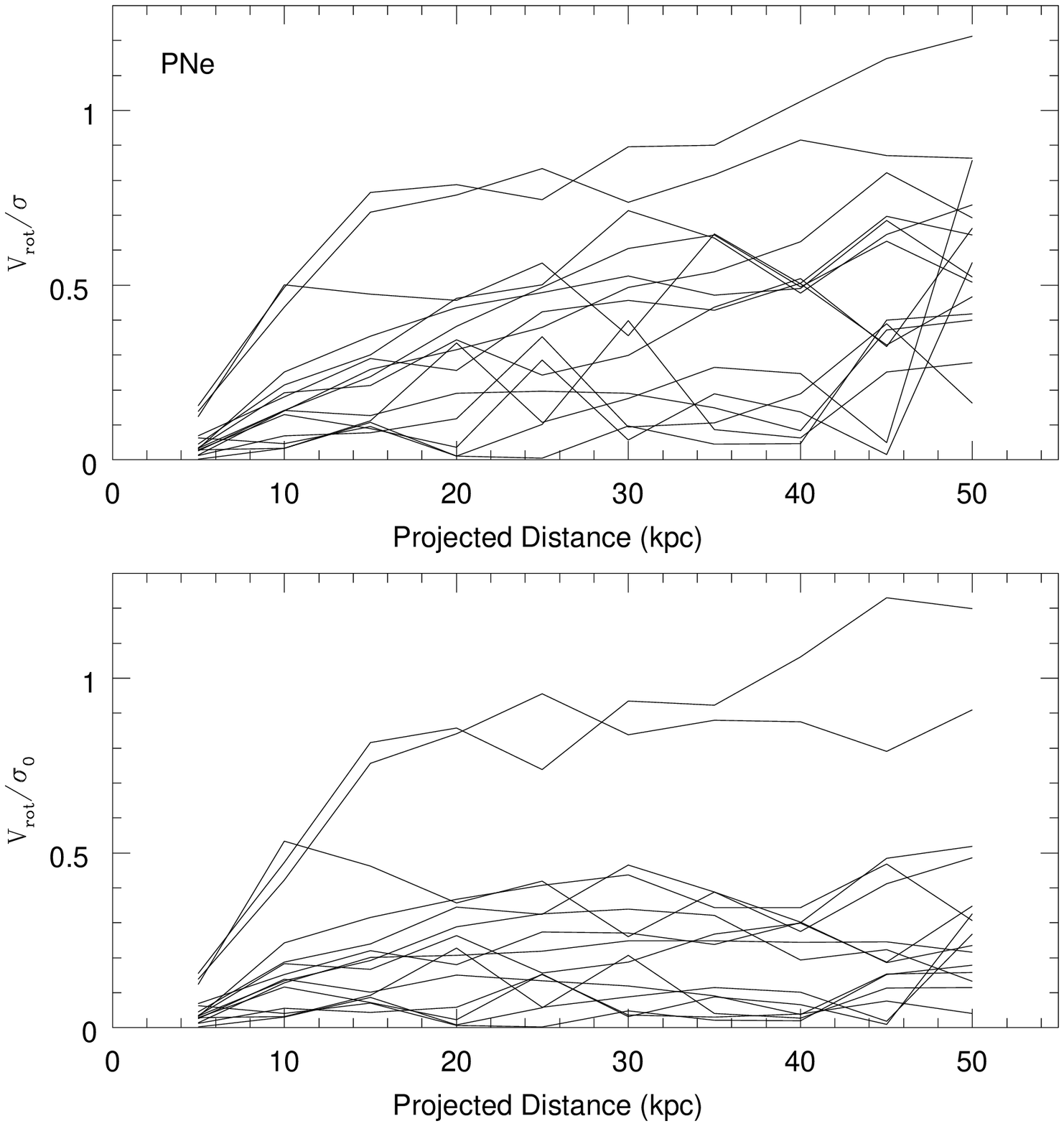,width=8.5cm}
\caption{ 
The same as Figure 9 but for the different 5 merger models
(M2, M3, M4, M5, and M6) viewed from three
different directions (i.e., projected onto
the $X$-$Y$,  $X$-$Z$, and  $Y$-$Z$ plane).
}
\label{Figure. 13}
\end{figure}

\begin{figure}
\psfig{file=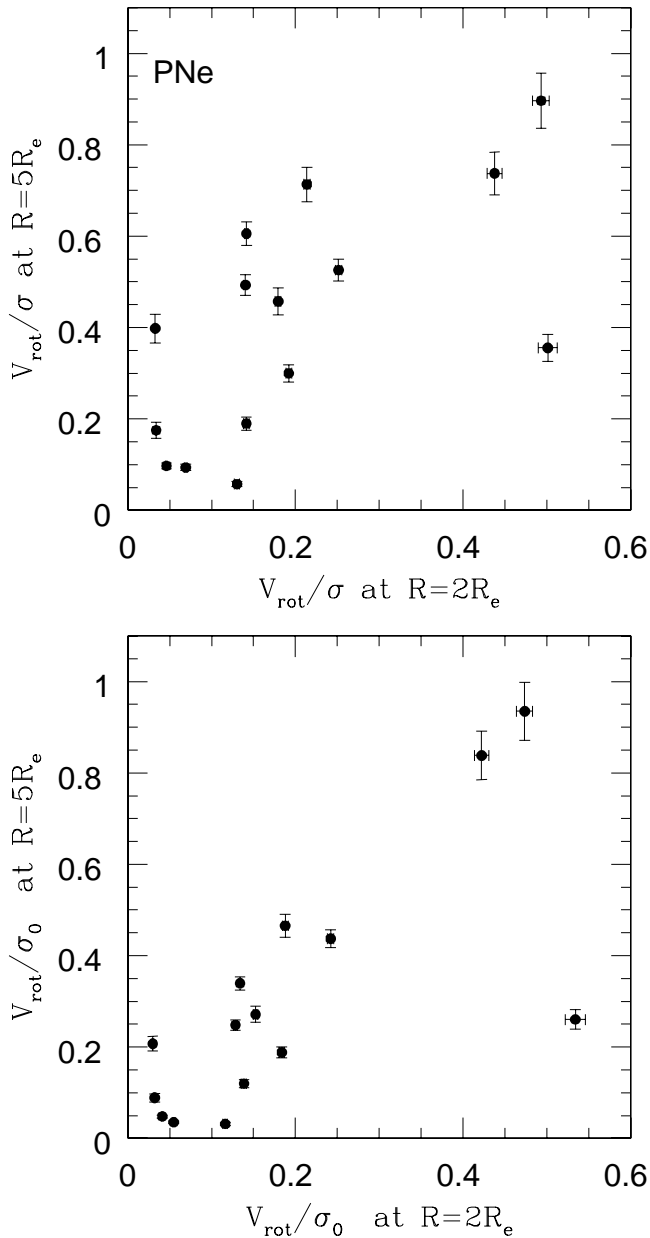,width=8.5cm}
\caption{ 
The same as Figure 10 but for the different 5 merger models
(M2, M3, M4, M5, and M6) viewed from three
different directions (i.e., projected onto
the $X$-$Y$,  $X$-$Z$, and  $Y$-$Z$ planes).
}
\label{Figure. 14}
\end{figure}

\subsection{Parameter dependences}

\subsubsection{Orbital configurations}
The dependences  of
dynamical properties of PNSs in elliptical galaxy formed by major galaxy
merging with $m_2$ = 1.0 on orbital configurations of the merging 
are described as follows. \\

(1) Irrespective of orbital configurations,
the radial profiles of the projected PNe number densities in PNSs 
can be fitted to the power-law profile for the entire halo regions
of their host elliptical galaxies.  The simulated PNSs of ellipticals in all models
in the present study have the mean PNe densities more than an order of magnitude  
higher than those of stellar halos of their merger progenitor spirals.
The central values  of the projected PNe number densities do not depend
on orbital configurations.
These results suggest that structural properties of outer PNSs (thus stellar halos)
can be significantly different between spirals and ellipticals. \\

(2) Most of PNSs can show a significant amount of rotation in the outer halo
regions ($R$ $>$ $5R_{\rm e}$) of their host elliptical galaxies,
though the maximum values of the rotational velocities ($V_{\rm rot}$) depend  on
the viewing angles of the ellipticals.
The 2D velocity fields of most PNSs shows minor-axis rotation 
in their halo regions, as seen in the main bodies 
($R$ $<$ $2R_{\rm e}$)  of elliptical galaxies
(e.g., Bendo \& Barnes 2000).
The details of the 2D velocity field of PNSs 
depend strongly on orbital configuration of galaxy merging. 
For example,  the PNS of the elliptical formed from a retrograde-retrograde  
merging (model M4) shows a more inhomogeneous and more complicated  2D velocity
field in its edge-on view than that from a prograde-prograde merging,
though there are no significant differences in the 2D fields between
the two for the face-on view (See Figures 11 and 12 and compares the two velocity
fields with each other).
The 2D velocity dispersion fields  does not so strongly depend on orbital 
configurations than the 2D velocity ones.\\

(3) $V_{\rm rot}/\sigma$ of PNSs are more likely to be higher in the outer   
halo regions of elliptical galaxies 
($R$ $>$ 25 kpc corresponding to $\simeq$ $5R_{\rm e}$) than in   
the central regions  $R$ $\sim$ 5 kpc,
as shown in Figure 13.
This result suggests that PNSs of elliptical galaxies
are likely to show rotational kinematics in their outer halo regions,
if they are formed from major galaxy merging.
$V_{\rm rot}/{\sigma}_{0}$ (rotation curve normalized to
the central velocity dispersion ${\sigma}_{0}$)
has the radial dependence  similar to
$V_{\rm rot}/\sigma$ (Figure 13). 
The flat rotation curve of $V_{\rm rot}/{\sigma}_{0}$ seen
in most of the models for their outer halo regions ($R$ $>$ 25 kpc )
suggests that such flattened shapes of $V_{\rm rot}/{\sigma}_{0}$
is one of generic trends of stellar halos kinematics in elliptical galaxies
formed from major merging. \\

(4) There is a weak correlation between 
$V_{\rm rot}/\sigma$ at $R$ = $2R_{\rm e}$ and
$V_{\rm rot}/\sigma$ at $R$ = $5R_{\rm e}$ in the sense
that PNSs with larger $V_{\rm rot}/\sigma$ at $R$ = $2R_{\rm e}$
are more likely to show larger $V_{\rm rot}/\sigma$ at $R$ = $5R_{\rm e}$
(See Figure 14).
This result suggests that kinematics of outer stellar halos 
in elliptical galaxies can correlate
with that of their main bodies,
if elliptical galaxies are formed from major merging. 
Similar trend can be seen for $V_{\rm rot}/{\sigma}_{0}$,
which implies that elliptical galaxies with the main bodies
rotating more rapidly are likely to show a larger amount of
rotation in their outer halo regions. 
The derived two correlations, though weak, can be used as theoretical predictions
that should be compared with future observations of PNe kinematics in
elliptical galaxies.

(5) These radial differences of $V_{\rm rot}/\sigma$ characteristics
of PNSs of elliptical galaxies formed by major merging
(See Figures 13 and 14) can be derived {\it only if we 
can obtain kinematical data for outer halo components ($R$ $\sim$ $5R_{\rm e}$)
of elliptical galaxies.} 
This implies that we need to investigate 
observationally and theoretically
kinematical properties of stellar halos beyond $2R_{\rm e}$
to give some  constraints on elliptical galaxy formation
and thus that PNe kinematical studies are ideal for this purpose.

\begin{figure}
\psfig{file=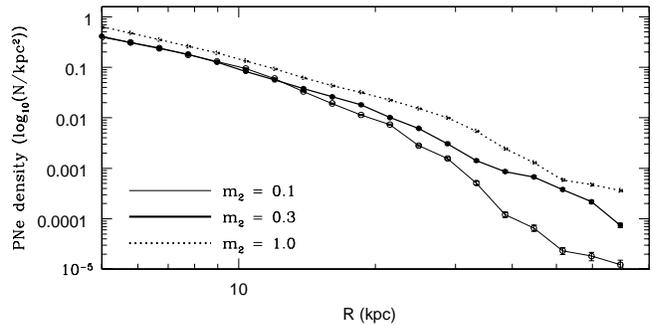,width=8.5cm}
\caption{ 
The same as Figure 4 but for the three  merger models
with different $m_{2}$ (M2, M7, and M8).
}
\label{Figure. 15}
\end{figure}

\begin{figure}
\psfig{file=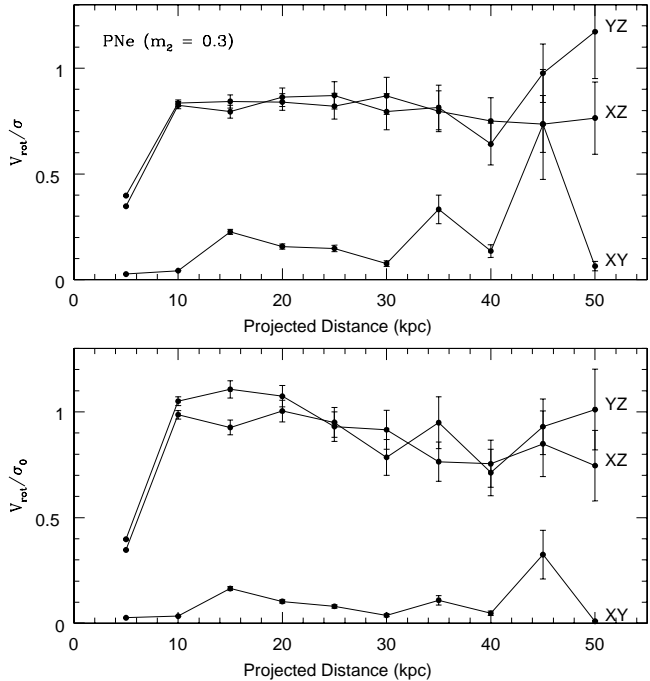,width=8.5cm}
\caption{ 
The radial profiles of $V_{\rm rot}/\sigma$ (upper)
and the normalized rotation curve
$V_{\rm rot}/{\sigma}_{0}$ (lower) for the unequal-mass merger model M8
with $m_{2}$ = 0.3
viewed from  three different viewing angles
(projected onto the $X$-$Y$,  $X$-$Z$, and  $Y$-$Z$ planes).
}
\label{Figure. 16}
\end{figure}

\subsubsection{Mass ratios ($m_2$)}

The dependences  of
dynamical properties of PNSs in merger remnants on $m_2$ 
are described as follows. \\

(1) The projected PNe number densities 
${\rho}_{\rm PNe}$ (${\log}_{10}$(N/kpc$^2$))
are globally lower in the remnants
with smaller $m_{2}$, in particular,
for the outer parts ($R$ $>$ $5R_{\rm e}$ corresponding to $\sim$ 25 kpc)
of PNSs (See Figure 15).
This means that more flattened elliptical galaxies are more likely to show
lower ${\rho}_{\rm PNe}$, because 
the morphological properties of the host galaxies formed from these mergers with lower
$m_{2}$ show more flattened intrinsic shapes. 
The power-law slope of the density profile 
${\rho}_{\rm PNe}$ (${\log}_{10}$(N/kpc$^2$) does not strongly depend
on $m_{2}$ for $R$ $\le$ $5R_{\rm e}$, however, there 
is a clear sign of steeper slope
in the model with $m_{2}$ = 0.1 for $R$ $>$ $5R_{\rm e}$
owing to the rather low PNe density there. 
The result implies that the ${\rho}_{\rm PNe}$ is steeper in Es formed from major
galaxy merging (e.g., $m_2$ $>$ 0.3) than in S0s from unequal-mass/minor galaxy
merging with ($0.1$ $<$  $m_2$ $<$ 0.3).

(2) The kinematics of PNSs in E/S0s formed from galaxy merging with
lower $m_2$ show more rapid rotation, larger maximum values of $V_{\rm rot}$,
and steeper radial gradients of $V_{\rm rot}$
for $R$ $<$ $2R_{\rm e}$ (See Figure 16). 
Given the fact that remnants of galaxy merging with smaller $m_2$
show more flattened shapes, this result suggests that 
more flattened E/S0 have higher $V_{\rm rot}$ and steeper radial $V_{\rm rot}$ gradients.
Irrespective of $m_2$,
the outer regions of PNSs ($R$ $\ge$ $5R_{\rm e}$) can show high $V_{\rm rot}$ 
and thus large $V_{\rm rot}/\sigma$, owing to the nearly flat rotation curve
and the radially decreasing $\sigma$ (See Figure 17).
This suggests that stellar halos
in E/S0s formed by galaxy merging show different kinematics compared with
that of the Galaxy with little rotation.

(3) The larger values  of  $V_{\rm rot}/\sigma$ in the PNS for 
edge-on projections (i.e., $X$-$Z$ and $Y$-$Z$)
in S0s formed from unequal-mass merging
suggests  that rotational terms in Jeans equations
should be considered when we estimate total masses of S0s 
using PNe data (See Figure 16).
The small values  of $V_{\rm rot}/\sigma$ for the face-on projection
(i.e., $X$-$Y$) imply that the total mass of an S0 (thus mass-to-light-ratio)
can be significantly underestimated if it is viewed from face-on
and if the mass is estimated by the Jeans equation with correction terms
of rotation
(and if the projected  central velocity dispersion 
does not depend on the viewing angle).
The observed small $M/L$ in some early-type galaxies
(e.g., in NGC 3379) could be due to
this viewing angle effect.

\begin{figure}
\psfig{file=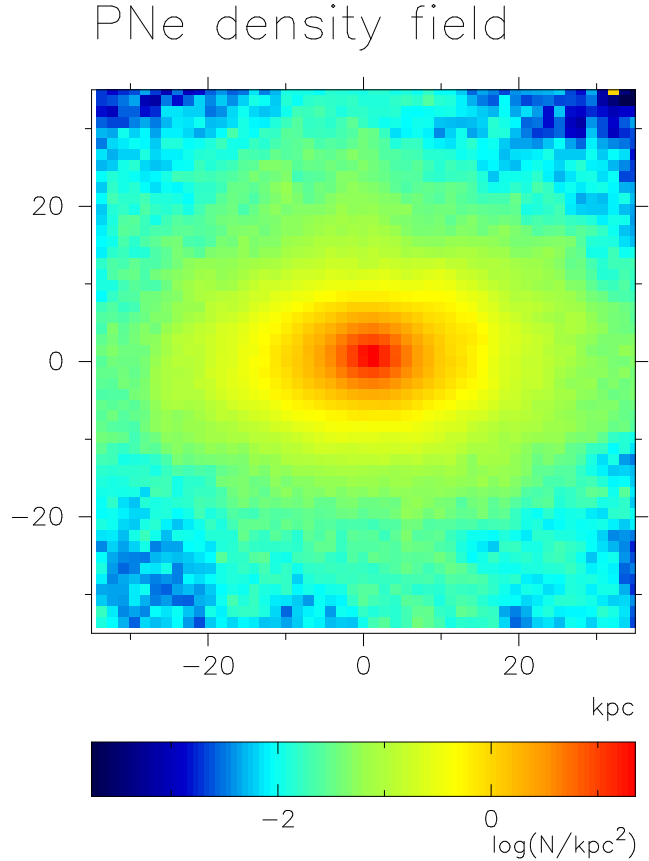,width=8.5cm}
\caption{ 
The same as Figure 1 but for the model M9.
${\alpha}_{\rm PNe}$ of  28.2
PNe per $L_{\odot}$ in $B$-band is adopted so
that the  PNe number density at the effective radius
of the merger remnant can be consistent with the observed
one for the NGC 5128 PNS.
}
\label{Figure. 17}
\end{figure}

\begin{figure}
\psfig{file=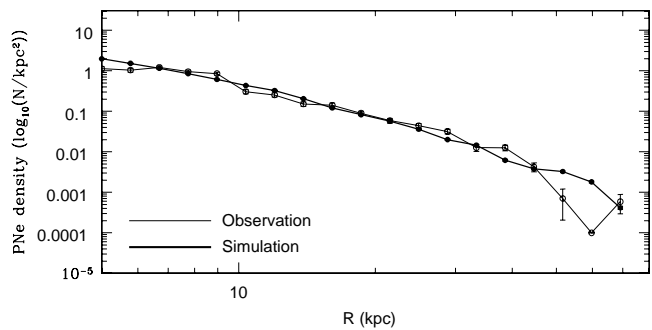,width=8.5cm}
\caption{ 
The same as Figure 3 but for
and the observational data (thin)
by PFF04a
and the model M9 (thick).
}
\label{Figure. 18}
\end{figure}

\begin{figure}
\psfig{file=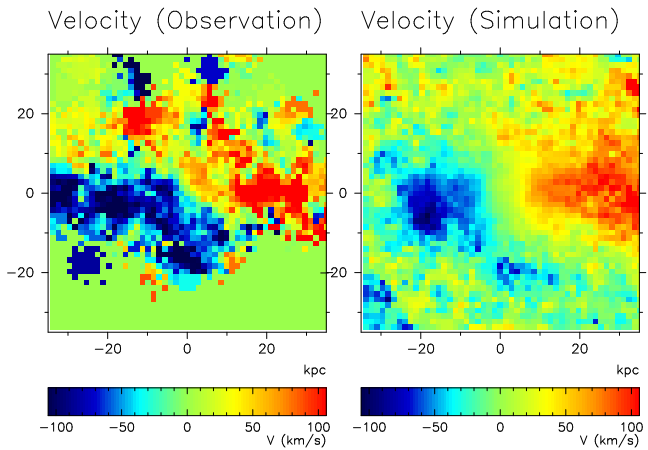,width=8.5cm}
\caption{ 
The same as Figure 5 but for the observational data (left)
and the model M9 (right).
}
\label{Figure. 19}
\end{figure}

\section{Comparison with observations of NGC 5128 PNe systems}

\subsection{Selection of the best model}

We present the results of one of the best models (model M9) in which
the observed structural and kinematical properties of the PNS in
NGC 5128 (PFF04a) are the most self-consistently reproduced. 
We select this best model among the 32 models investigated in the present study
by checking whether the following five fundamental observational  results
(e.g., Israel 1998; PFF04a) can be reproduced reasonably well in each model:
(1) $R_{\rm e}$ = 5.2 kpc, $M_{\rm l}$ (total luminous mass) 
= 1.4 $\times$ $10^{11}$ $M_{\odot}$ (for $M_{\rm l}/L_{\rm B}$ = 3.5), 
and spherical appearance of the elliptical galaxy,
(2) the projected PNe number density ${\rho}_{\rm PNe}$ (${\log}_{10}$(N/kpc$^2$)
$\simeq$ 1.0 at $2R_{\rm e}$ and the slope of  the power-law profile similar to $-2.5$,
(3) the  rotation curve that rises till $R$ $\simeq$ $2R_{\rm e}$
and becomes flat at $R$ $>$ $2R_{\rm e}$ with the maximum
$V_{\rm rot}$ of $\simeq$ 100 $\pm$ 20  km s$^{-1}$,
(4) the central velocity dispersion ${\sigma}_{0}$ of $\simeq$ 140 $\pm$ 10 km s$^{-1}$, 
and (5) the 2D velocity field  showing 
both minor and major axes rotation with the zero velocity
curve/contour  (ZVC) twisting significantly.  
 
Owing to the purely collisionless nature of the present simulations,
we can match any model to the above observation (1)
by rescaling the size and the mass of the simulation
Therefore, the above (2) - (5) constraints  can be used in selecting 
the best possible model in the present study.
Major merger models  with $m_{2}$ = 1.0 can explain the above (2), (4), and (5)
reasonable well, however they have difficulties in explaining (3) owing to
slowly rising $V_{\rm rot}$ with smaller maximum values of $V_{\rm rot}$.
Unequal-mass merger models  with $m_{2}$ = $0.1-0.3$ can become early-type E/S0s,
however,  they are very flattened in shapes,  strongly supported by rotation,
and have 2D velocity fields that are not similar to the observed ones.
These models accordingly  can not explain the above (1), (4), and (5) 
in {\it a fully self-consistent manner} and thus can not be regarded as
the best model. 
Thus one of the best models can be unequal-mass mergers with 
$m_{2}$ of somewhere between  0.3 and 1.0  (viewed from a certain direction): However,
it should be stressed here that we possibly could miss out the major merger
model with $m_{2}$  = 1.0 that can explain the above five points self-consistently,
owing to the limited number of the simulated models.

The best model for which we show the results below
is the model 9 with $m_{2}$ = 0.5 in which a highly inclined
orbital configuration and a smaller pericentre distance are assumed. 
We investigate dynamical properties of the PNS in this model
at $T$ = 4.5 Gyr when the remnant is dynamically well relaxed
to show regular distributions of PNe both in the 2D distribution 
of the projected number density and in the radial one (See Figures 17 and 18).
This model with smaller pericenter distance and a highly inclined disk 
with respect to the orbital plane can 
form stellar shells and gaseous rings perpendicular to the major axis
of the merger remnant if gaseous dissipation is included (Bekki 1997, 1998a). 
Thus, our future more sophisticated model with gaseous dissipation and
star formation will be capable of explaining the observed HI distribution 
perpendicular to the photometric major axis and the outer
shells as well as the above five observations. 
Dynamical properties of globular cluster systems and fine structures
observed in NGC 5128 (Peng et al. 2002; 2004b, c) will be discussed
in the best  model(s) of our future studies (Bekki \& Peng 2005, in preparation). 

\begin{figure}
\psfig{file=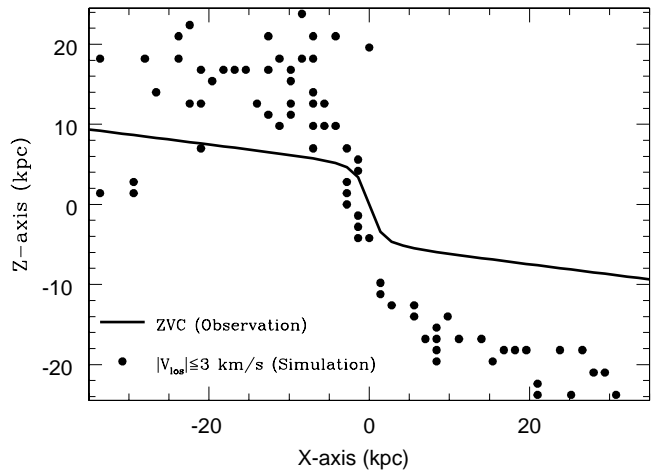,width=8.5cm}
\caption{ 
A comparison between the observed ZVC (zero velocity curve) and
the simulated one for the model M9.
The observed one (PFF04a) is represented by a solid line and
the filled circles indicates the location of cells with
$|V_{\rm los}|$ (the absolute magnitude of
line-of-sight-velocities)  $\leq$ 3.0 km s$^{-1}$.
The locations of the cells with $X$ $<$ 0 kpc and
$Z$ $<$ 0 kpc and those with $X$ $>$ $0$ kpc
and $Z$ $>$ 0 kpc are not plotted for the simulation
so that we can more readily infer the possible ZVC of the simulation model
from the distribution of the cells with $|V_{\rm los}|$ $\leq$ 3.0 km s$^{-1}$.
The mathematical expression of the observed ZVC is given in the main text.
}
\label{Figure. 20}
\end{figure}

\begin{figure}
\psfig{file=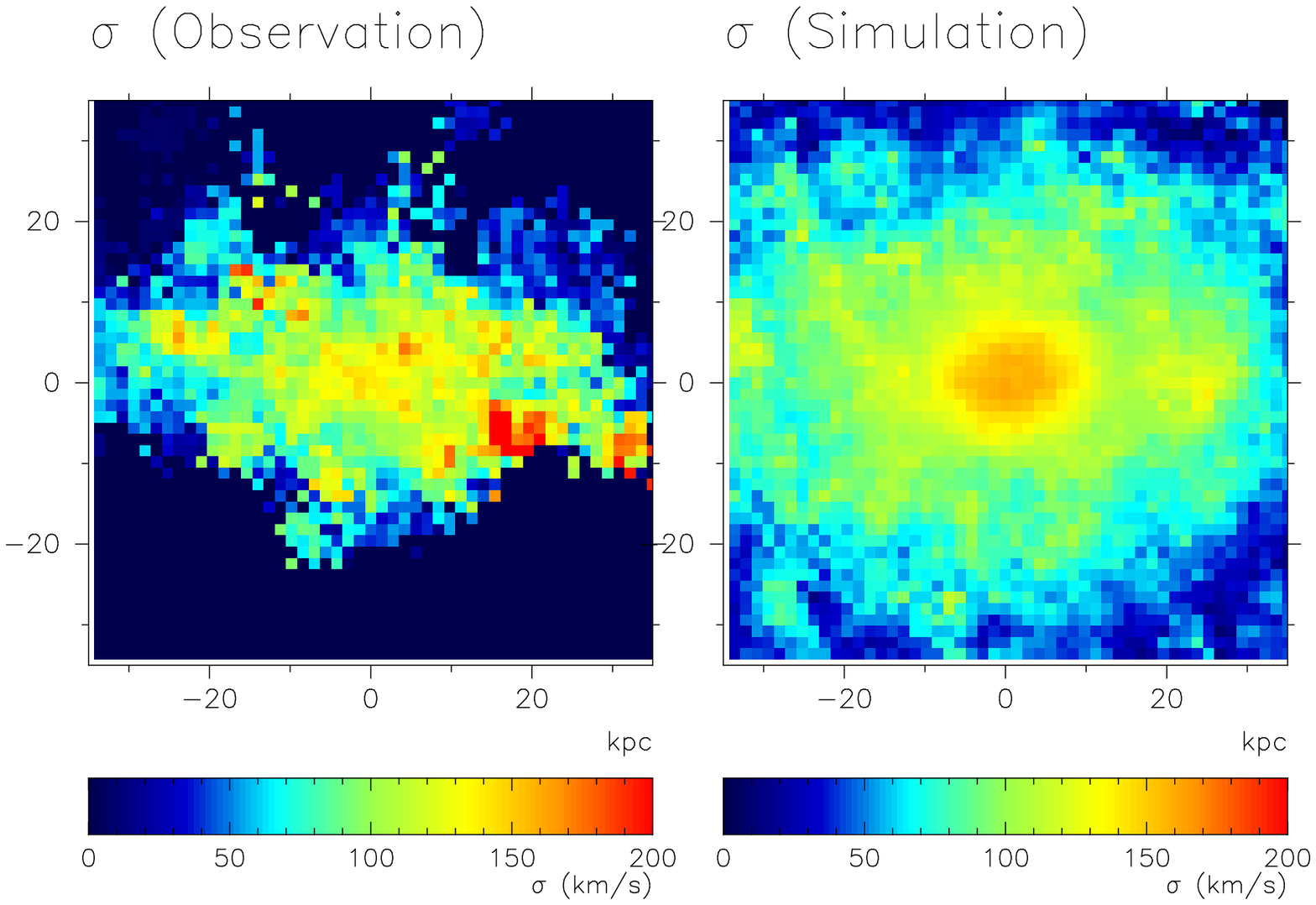,width=8.5cm}
\caption{ 
The same as Figure 6 but for the observational data (left)
and the model M9 (right).
}
\label{Figure. 21}
\end{figure}

\begin{figure}
\psfig{file=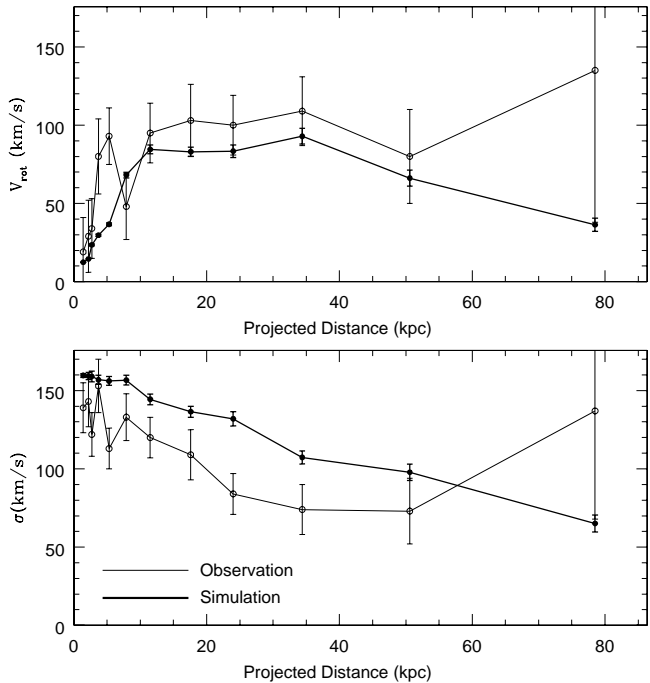,width=8.5cm}
\caption{ 
The same as Figure 7 but for the observational data (thin)
and for the model M9 (thick).
In order to compare the simulation result with the observations
in a more self-consistent manner, the way of binning is
exactly the same between the simulation and the observation.
}
\label{Figure. 22}
\end{figure}

\subsection{Advantages and disadvantages of the best model}

Figures 19 presents the comparison between the simulated 2D velocity field of
the PNS of the best model and the observed one. 
In order to match the observation with the simulation,
the observed coordinate $-X$ ($Y$) for the PNS in NGC 5128 (PFF04a) 
is set to represent $+X$ ($Z$) in Figure 19 (and 20, 21) for convenience.
The best model is consistent with the observation at least qualitatively
in the sense that it shows (1) strong major-axis rotation extending to 35 kpc 
($\simeq$ $7R_{\rm e}$), (2) weaker but significant rotation along the minor 
axis, and (3) the line of zero velocity (referred to as zero velocity curve,
ZVC) both misaligned and twisted with respect
to the major axis of the PNe distribution.

Figure 20 more clearly describes  how the ZVC looks like in the 2D velocity field
and whether it is consistent quantitatively with the observed ZVC. 
The observed ZVC can be parameterized as follows (PFF04a):
\begin{equation}
Z = \frac{-X(0.125|X|+5)}{\sqrt{(X^2+2.56)}}
\end{equation}
where $X$ and $Z$ are exactly the same as the coordinate $X$ and $Z$ in
the simulations.  As shown in Figure 20,
the simulated ZVC starts twisting at $Z$ $\sim$ $\pm$ 10 kpc whereas the observed
ZVC starts twisting at $Z$ $\sim$ $\pm$ 4 kpc.
The direction of the simulated ZVC for $|X|$ $>$ 10 kpc is broadly
consistent with the observed one, though the simulated ZVC is more 
aligned with the minor axis of the 2D PNe  distribution (shown in Figure 19) 
compared with the observed one for $|X|$ $\le$ 10 kpc.
These less successful reproduction of the model suggests that  
(1) some physics which are not included in the present study
(e.g., gaseous dissipation and star formation) 
should be considered to reproduce fully self-consistently
the observations
and (2) we need to explore wider sets of model parameters 
(e.g., merger orbits)  to
find a fully self-consistent model.

Figure 21 presents the comparison between the simulated 2D velocity
dispersion field of the PNS and the observed one (PFF04a).
The simulation is consistent broadly with the observation in the sense
that (1) it shows an inner flattened shape in the 2D velocity dispersion field  
and  (2) radial gradient of velocity dispersion along the minor axis
is steeper than that along the major axis.
Interestingly, the observation shows an isolated region with high 
velocity dispersion ($\sim$ 200 km s$^{-1}$) around ($X$, $Z$) = (15, $-5$) (kpc).
An apparently isolated  region with moderately high velocity dispersion ($\sim$ 120 km s$^{-1}$)
can be discernibly seen also in the simulation around ($X$, $Z$) = (25, $-2$) (kpc).
Although it is not so clear whether 
the presence of such local, dynamically hot regions 
in the 2D velocity dispersion fields have some physical meanings
of elliptical galaxy formation,
more extensive comparison between observations and simulations 
in terms of the locations of the isolated, dynamically hot regions 
should be made to determine the best model
after a larger set of PNe kinematical data become available, in particular,
for the regions with $X$ $>$ 0 kpc and $Z$ $<$ $-10$ kpc.

Figure 22 demonstrates that the best model can explain 
the observed rotation curve ($V_{\rm rot}$) profile  reasonably well
for $R<3$ kpc and $10 <R 50$ kpc.
However there is a significant difference in $V_{\rm rot}$ profiles
between the observation and the simulation for $R \sim$ 5 kpc:
The observational rotation curve rises more rapidly than
the simulated one for the inner region of NGC 5128. 
The radial profile of velocity dispersion 
($\sigma$) in the model is systematically
higher (a factor of 1.2 at $R$ $\simeq$ $2R_{\rm e}$) than the observed one.
We suggest that 
these less successful reproduction of the best model 
for $V_{\rm rot}$ around $R=5$  kpc and the $\sigma$ profile
is due to the model's
not including PNe formation from gas: 
All PNe are assumed to originate  from collisionless old stellar
disks. We can expect that if PNe can form from gas, new PNe 
have a larger amount of rotation and smaller amount of random kinetic energy
owing to efficient energy  dissipation of random kinematic energy
during galaxy merging. 
We thus suggest that our future more sophisticated model with gaseous 
dissipation can more successfully reproduce the observed radial profiles
of $V_{\rm rot}$ and $\sigma$.

\begin{table*}
\centering
\caption{Mass comparison}
\begin{tabular}{cccc}
Model no. & 
Fitted mass  ($M_{\rm T,F}$) & 
Actual mass ($M_{\rm T,A}$) & 
$M_{\rm T,F}/M_{\rm T,A}$  \\
M1 (edge-on) &  $1.7 \times 10^{11} {\rm M}_{\odot}$ &  
$8.6 \times 10^{11}  {\rm M}_{\odot}$   &   0.20 \\
M1 (face-on) &  $7.4 \times 10^{10} {\rm M}_{\odot}$ &  
$8.6 \times 10^{11}  {\rm M}_{\odot}$   &   0.09 \\
M8 (edge-on) &  $9.3 \times 10^{11} {\rm M}_{\odot}$ & 
$7.1 \times 10^{11}  {\rm M}_{\odot}$   &   1.31 \\
M8 (face-on) &  $1.5 \times 10^{10} {\rm M}_{\odot}$ &  
$7.1 \times 10^{11}  {\rm M}_{\odot}$   &   0.02 \\
\end{tabular}
\end{table*}

\section{Discussion}

\subsection{Constraints on galaxy formation from stellar halo kinematics}

PFF04a have revealed that the PNS of NGC 5128 has a significant
amount of rotation with $V/\sigma$ between 1 to 1.5
in its outer stellar halo region ($R$ $>$ $5R_{\rm e}$).
The present study has  shown that outer stellar halos
in most of major merger models have  a significant amount of rotation
and thus suggested that a rotating stellar halo seen in NGC 5128
is not just a special case but a rather general trend  of elliptical
galaxies, if ellipticals are formed by galaxy merging. 
The kinematics of PNSs in several elliptical galaxies
have been investigated so far 
(Ciardullo et al. 1993; Hui et al. 1995; Arnaboldi et al. 1998;
Mendez et al. 2001; Romanowsky et al. 2003; Napolitano et al. 2004;
PFF04a) and found to show rotation in some of the ellipticals.
Although these observational studies help the authors to
provide a reasonable dynamical model (e.g., total mass of a galaxy) 
for each individual case, 
the total number of PNSs investigated is too small for them to
make any robust conclusions on the general trend of kinematics
in {\it outer stellar halos} ($R$ $>$ $5R_{\rm e}$) of elliptical galaxies. 

We suggest that extensive systematic studies of kinematical
studies of PNe for $2R_{\rm e}$ $\le$ $R$ $\le$ $10R_{\rm e}$ 
for a larger number of elliptical galaxies
are doubtlessly worthwhile,
because  the kinematics of outer stellar halos 
can  provide strong constraints
on elliptical galaxy formation models,
such as the monolithic collapse scenario (e.g., Larson 1974; Carlberg 1984)
the merger one (e.g., Toomre 1977; Barnes 1992),
and multiple merging of subgalactic clumps or dwarfs 
(e.g., C\^ote et al. 2002). 
Although kinematical properties of outer stellar halo have not been investigated 
by the monolithic collapse scenario (Larson 1974),
it is unlikely that the outer stellar halos in elliptical galaxies
formed in monolithic collapse
show more rapid  rotation than the inner main bodies,
because random kinematical energy is likely to be
dissipated away more efficiently 
to form a stellar system
with rotation in the inner regions of protogalaxies  where 
gas densities are higher (thus more dissipation is highly likely). 
If elliptical galaxies are built up through multiple merging of dwarfs 
with random orientation of  merging, 
there appears to be no strong physical reasons for them
to show a significant amount of rotation in their outer stellar halos.
Thus stellar halo kinematics that can be proved by kinematical
studies of PNSs in currently ongoing observations
can be used to assess the viability of each of
the above three scenarios of elliptical galaxy formation.

\subsection{Halo-host connection}

The present study discovered that both  the radial profiles
of projected PNe number densities (${\rho}_{\rm PNe}$)  
and the kinematical properties of PNSs in
merger remnants depend strongly
on the mass ratios of the  merger progenitor spirals 
(See Figures 14, 15,  and 16). 
Given  the fact that morphological properties of merger
remnants are  demonstrated by numerical studies to depend strongly on
the mass ratios (e.g., Bekki 1998b; Naab \& Burkert 2003), 
the above discovery can provide the following three
predictions on the possible correlations in
dynamical properties  between PNSs and
their host galaxies.
Firstly,  more flattened E/S0 galaxies are more likely to 
have  smaller mean
${\rho}_{\rm PNe}$ and higher $V_{\rm rot}/\sigma$ at $5R_{\rm e}$,
because unequal-mass galaxy mergers with $m_2$ $\simeq$ $0.3-0.5$ 
finally become more flattened E/S0s (Bekki 1998b; Naab \& Burkert 2003).
Secondly, stellar halos of S0s with thick disks are more likely to show
a significantly smaller ${\rho}_{\rm PNe}$ compared with those
of Es,  because S0s with thick disks can be formed from
lower mass-ratio unequal-mass 
galaxy mergers ($m_{\rm 2}$ $\simeq$ $0.1-0.3$). 
Thirdly, E/S0 galaxies can show a correlation 
in kinematics between $V/\sigma$ (e.g., at $R$ $\simeq$ $2R_{\rm e}$)
of the main bodies and that of their stellar halos
(e.g., at $R$ $\simeq$ $5R_{\rm e}$),
if most of them are formed from galaxy merging.

These three predictions will be more readily tested in future and ongoing
kinematical studies of main bodies and stellar halos of E/S0 
galaxies by the planetary nebula spectrograph (Douglas et al. 2002)
and integrated field spectrographs like SAURON (Bacon et al. 2001).
Furthermore, as have been demonstrated by HH00 an PFF04a,
both the MDF and the kinematics of the stellar halo in NGC 5128 can be 
significantly different 
from those of the stellar halo of the Galaxy in the sense that
the stellar halo of NGC 5128 are more metal-rich and more strongly
rotating. These differences in stellar halo properties may well
be remarkable and fundamental 
differences between spiral and elliptical galaxies,
though extensive observational studies of PNSs in spirals beyond the Local Group
that can prove dynamical properties of the stellar halos
have not been yet conducted. 
Although the MDF  was investigated for the stellar halo
of the edge-on S0 NGC 3115 
(Elson 1997; Kundu \& Whitmore 1998), 
systematic studies on the MDFs of stellar halos
and dynamical properties of PNSs
in S0s have not been done yet.
Thus future observational studies of structural and kinematical
properties of PNSs in spirals and S0s,
combined with those ongoing for ellipticals,
will provide a new clue to the origin of the Hubble sequence.

\subsection{Mass Estimates}

One ongoing field of study using PNSs is the total mass of elliptical
galaxies.  Typically, analysis of this sort use the integrated
rotation and velocity dispersion profiles of the PNe to derive a total
mass within a given radius (e.g.\ PFF04a, Romanowsky et al.\ 2003,
M\'endez et al.\ 2001).  As one can see from Figure~9, a single galaxy
can have a wide range of rotation and dispersion profiles depending on
the viewing angle, and this could plausibly lead to biases in the mass
estimation.  We test this by adopting the mass estimation procedure
outlined in PFF04a, which uses the spherical Jeans equation and
assumes an isotropic distribution of orbits.  While this is a simple
assumption, it is one that is typically used and there has yet to be
any strong kinematic evidence that orbits in ellipticals are
strongly anisotropic.  When we fit the kinematic profiles of the major
merger model from the face-on and edge-on views, we find that the
face-on view gives a mass that is a factor of two lower than the
edge-on view.  The reason for this can be seen qualitatively in
Figures 7 and 8 which shows that the rotation and velocity dispersion
profiles are both lower for the face-one view.  

Even more intriguing, the simulations consistently show that the dark
matter is at a higher velocity dispersion than the stars.  In fact,
while the face-on view under-represents the mass relative to the
edge-on view, {\it both} viewing angles underestimate the total mass.
In order to estimate the total mass of a merger remnant from
PNe kinematics and thereby discuss this point in a more
quantitatively, we use the same method as those used by Hui et al. (1995)
and PFF04a:  We apply the spherical Jeans equation to
the major-axis rotation and line-of-sight velocity dispersion
profile (e.g., Figures 7 and 8) in a merger remnant (i.e.,
an elliptical galaxy) to derive the dynamical mass
of the remnant (See PFF04a for more details on the methods of 
mass estimation from PNe kinematics).

Table 2 summarizes  the actual masses of the simulated E/S0s
within 45 kpc ($M_{\rm T,A}$) and the masses estimated from
the simulated radial velocity dispersion profiles and the Jeans
equation used in Peng et al. (2004a) within 45 kpc 
($M_{\rm T,F}$)  for two interesting cases.
Compared to the total amount of mass in the simulation, our simple
mass estimation from the stellar data viewed edge-on (best case) underestimates
the total mass within 45~kpc by a factor of 5.  In one model for a
minor/unequal-mass  merger viewed face-on, 
the total mass is underestimated by a
factor of 50 and requires no dark halo.  If ellipticals result
from these types of mergers, then this could be a very important
consideration, especially considering that current mass estimates for
ellipticals are surprisingly low.  This matter warrants further
investigation with a variety of different elliptical formation models.
(Dekel et al.\ 2005 have recently discussed this problem with a
complementary approach and also find masses can be underestimated).
However, we point out that the supposedly low $M/L$ recently measured
are for intermediate-luminosity galaxies (e.g.\ NGC~3379 and NGC~5128)
and not for all galaxies.  If this ``stellar kinematic bias'' is 
the reason for anomalously low mass estimates, then it may be more
pronounced in lower luminosity ellipticals, providing a clue to their
formation histories.

\section{Conclusion}

We have numerically investigated structural and kinematical properties
of PNSs in elliptical galaxies formed by galaxy merging 
in order to elucidate the origin of the observed physical properties
of PNSs.
We have mainly investigated the radial profiles of projected number densities,
rotational velocities, and velocity dispersion of PNSs and
the two-dimensional velocity fields of PNSs in elliptical galaxies.
We also compared the simulated dynamical properties of PNSs with
the corresponding observations of the PNS in NGC 5128
and thereby tried to provide the best merger model for the PNS in NGC 5128. 
We summarize our principal results as follows.

(1) The radial densities profiles of PNSs  can be fitted to the power-law 
in the entire halo regions
of elliptical galaxies formed from major merging ($m_{2}$ = 1.0). 
The projected PNe number densities (${\rho}_{\rm PNe}$)
at $R$ $\sim$ 5$R_{\rm e}$
in elliptical galaxies are more than an order of magnitude higher than
those of the halos of the merger progenitor spirals. 
These results do not depend strongly on the model parameters of major merging.
The main reason for  rather high ${\rho}_{\rm PNe}$
of PNSs is that a significant fraction ($\sim$ 10\%) of disk stars
are stripped from the disks and redistributed in the halo regions
during major merging.
These results suggest that elliptical galaxies 
have higher PNe number densities (${\rho}_{\rm PNe}$)
in their halos than spiral ones. \\

(2) PNSs of elliptical galaxies formed from major merging can show
a significant amount of rotation ($V_{\rm rot}/\sigma$ $>$ 0.5) in their
outer halo regions ($R$ $>$ $5R_{\rm e}$). 
The derived rotation in the outer halos results from 
the angular momentum redistribution of disk stars
during galaxy merging (i.e., conversion of orbital angular momentum
of merging two spirals into the intrinsic one of the merger remnant).
$V_{\rm rot}/\sigma$ of PNSs in ellipticals 
are more likely to be larger in the outer parts
than in the inner ones,  though radial profiles of $V_{\rm rot}/\sigma$
are diverse between different models.
$V_{\rm rot}/\sigma$ at $5R_{\rm e}$ can weakly correlate with
$V_{\rm rot}/\sigma$ at $2R_{\rm e}$ for the PNSs in such a way
that PNSs with larger $V_{\rm rot}/\sigma$ at $2R_{\rm e}$ are
likely to show larger $V_{\rm rot}/\sigma$ at $5R_{\rm e}$. 
This result implies that the kinematics of outer stellar halos
in elliptical galaxies can correlate with that of the main bodies.\\

(3) The two-dimensional (2D) fields of velocity and velocity dispersion 
of PNSs in elliptical galaxies formed by major merging
are  quite diverse depending on the orbital configurations of galaxy merging,
the mass ratios of the progenitor spirals, and the viewing angles of the galaxies.
For example, PNSs of ellipticals formed from prograde-prograde merging
can show more rapid rotation along the major axes of the PNe density profiles
in their 2D velocity fields compared with 
the retrograde-retrograde ones.
For most of the models, the 2D velocity fields show minor axis rotation 
in the outer halo regions of elliptical galaxies. 
It is doubtlessly worthwhile for future observational studies
to systematically investigate dynamical properties of PNS for a larger number of
elliptical galaxies 
and thereby to confirm whether the predicted uniformity and diversity
in  the 2D velocity fields of PNSs can be seen in elliptical galaxies. \\

(4) ${\rho}_{\rm PN}$ of PNSs are  more likely to be lower in 
E/S0s formed from galaxy merging with smaller mass ratios
(e.g., $m_2$ = 0.3, i.e.,  unequal-mass merging).
Furthermore, PNSs in the remnants of mergers with smaller $m_2$
have more flattened shapes and a larger amount of rotation ($V_{\rm rot}$)
both in their main bodies and in their outer halos.
Therefore these  results suggest that 
more flattened E/S0s are more likely to show
both lower ${\rho}_{\rm PNe}$ and higher $V_{\rm rot}$ (for a given
luminosity range of E/S0s). 
This predicted correlation can  be readily confirmed in future observations
of PNSs.\\

(5) The observed kinematical properties of the PNS in NGC 5128 
(e.g.,  $V_{\rm rot}/\sigma$ between 1 and 1.5  and minor axis rotation) can be 
broadly consistent with the present best
model with $m_2$ = 0.5, a small impact parameter,
and highly inclined initial disks (similar to a  polar-orbit).
The observed ``kink'' in the zero velocity curve (ZVC)
of the 2D velocity field of the PNS in NGC 5128
can be reproduced reasonably well, though
the location where ZVC begins to twist is different between
the best simulation model and the observation.
Some disadvantages of the present best model in explaining self-consistently
the observed kinematics of the PNS  
suggest that gas dynamics and star formation 
may well play an important role  
in the formation of the PNS observed in NGC 5128.\\
 
(6) The mass estimates of merger remnants viewed face-on are likely to
    be a factor of two lower than those viewed edge-on.  However, even
    the mass estimates of systems viewed edge-on can be low by a
    factor of 5, and in the worst case (face-on E/S0 model) can be low by a
    factor of 50.  If this conclusion is applicable to real
    early-type galaxies, it has interesting consequences for current
    observational work.

\section*{Acknowledgments}
We are  grateful to the referee for valuable comments,
which contribute to improve the present paper.
KB  acknowledges the financial support of the Australian Research 
Council throughout the course of this work.
The numerical simulations reported here were carried out on GRAPE
systems kindly made available by the Astronomical Data Analysis
Center (ADAC) at National Astronomical Observatory of Japan (NAOJ).
E. W. P. acknowledges support from NSF grant AST 00-98566.

\appendix

\section{Derivation of 2D  density and velocity fields}

In order to compare the simulated 2D density and velocity fields
with the observed ones  in a more self-consistent
manner, we adopt the methods that are quite similar to those used
for kinematical analysis of observational data of PNSs in elliptical 
galaxies (e.g., PFF04a).
We first determine the position angle (${\theta}_{\rm p}$)
of the major axis of the stellar mass distribution
projected onto the $x$-$z$ plane 
(i.e., perpendicular to orbital plane of  galaxy merging)
for each  model at final time step.
Then we rotate the model by $-{\theta}_{\rm p}$ within the $x$-$z$ plane
so that the major axis of the mass distribution can coincide with  the $x$-axis.
For convenience, the coordinate ($x$, $y$, $z$) is renamed as ($X$, $Y$, $Z$) 
after the rotation of the model. 
We represent the kinematics of a model
by estimating line-of-sight (LOS) velocity moments as a function of
position with a nonparametric smoothing algorithm.
At the position of each stellar particle,  we apply a local linear smoother
using a Gaussian kernel function with the smoothing length of 0.17 in our
units (corresponding to 3 kpc). 
We choose this value of 3 kpc, firstly because recent observational
study (PFF04a) chose this value and secondly because we can more clearly
see the detail of the 2D velocity field of a model throughout the outer
halo region (extending to $R$ $\sim$ $10R_{\rm e}$ without losing
resolution within $R_{\rm e}$.

We divide the entire halo region with the size of $4R_{\rm d}$
(= 70 kpc corresponding to $\sim$ 14$R_{\rm e}$) into 50 $\times$ 50 cells
for a merger model in each projection and estimate 
line-of-sight-velocity of $v(X_{i},Z_{i})$ for the $X-Z$ projection
($v(X_{i},Y_{i})$ and $v(Y_{i},Z_{i})$, for  
$(X_{i},Y_{i})$ and  $(Y_{i},Z_{i})$, respectively),
at the center of each cell, i.e.,  $(X_{i},Z_{i})$ ($i$ = $1-50$)
based on the smoothed velocity filed (described above).
Total number of cells in each projection
is fixed at 2500 for all merger models in the present study, 
because we can more clearly see the global changes of 2D velocity field
without suffering small-scale, and rapid variation resulting from small number of
particles in each cell for this cell number.
We mainly show the 2D velocity fields projected onto the $X$-$Y$ plane 
(referred to as the face-on view for convenience) and 
the $X$-$Z$ plane (the edge-on plane).
The same method is used for estimation the 2D density field of  
the PNS in model.

In converting the 2D stellar mass density field into the 2D PNe one,
we assume that the luminosity-specific PNe (number) 
density represented as ${\alpha}_{\rm PNe}$ is 9.4$\times 10^{-9}$
  PNe/$L_{\odot}$
(in $B$-band) 
This  represents the number of PNe expected in the
brightest 2.5~magnitudes of the PNLF, which is also roughly equivalent to
the typical PN survey depth.
Observations of M31's PNS (Ciardullo et al. 1989)
showed that ${\alpha}_{\rm PNe}$ can range 
from 2.9$\times 10^{-9}$to 39.3$\times 10^{-9}$,
and accordingly the adopted value above is consistent with these observations.
This value of 9.4$\times 10^{-9}$, which originates from the 2nd column of the table 4
in Ciardullo et al (1989),
is used for ${\alpha}_{\rm PNe}$ throughout this paper unless specified. 
While this is very obviously a simplification---$\alpha$ is known to
vary as function of metallicity, and observational incompleteness will
often vary across a galaxy---it is a good first-order
approximation to the true PN population for our purposes.
We also assume that  $M/L_{\rm B}=3.5$ for stellar components of galaxies
in all models.

The total number of PNe ($N_{\rm PNe}$) within $5R_{\rm e}$ of the
simulated elliptical galaxies is typically $\sim$ 350. 
As shown in Figure 1,
even for the small number of star particles (1000) in the halo of the disk,
the initial spherical distribution of PNe (stellar halo) 
can be reproduced reasonably well.
In most of the merger models,  typically 10 \% of the total stellar mass 
in the merger remnant can distribute throughout the halo regions ($R$ $>$ $2R_{\rm e}$).
Therefore, more than $10^4$ stellar particles (up to 5 $\times$ $10^4$) can
be used to derive the smoothed density (velocity) field for a model.
Thus we can derive the smoothly changing  density (velocity) fields
throughout the entire halo regions
from original,  more discrete density (velocity) ones 
for the adopted total number of stellar particles in simulations
by using the smoothing method described above.

We also derive the radial profiles of rotational velocity ($V_{\rm rot}$)
and velocity dispersion ($\sigma$) for each model
based on the same method as used for
observational analysis on the NGC 5128 PNS (PFF04a).
For example, when we try to derive the radial profile of 
$V_{\rm rot}$ along the major axis of a PNS , we use 
PNe within either a perpendicular distance of $\pm$ 2 kpc
from the major axis or a $\pm 10^{\circ}$ cone centered on the major axis.
This method is adopted 
so that more PNe 
can be included in the analysis of the radial profiles
for the outer halo regions,
where less number of stellar particles can prevent
us from making a reasonable estimate of  $V_{\rm rot}$ and $\sigma$. 
The errors in $V_{\rm rot}$ ($\sigma$) shown in figures of the
present paper (e.g., Figure 7)
are equal to $V_{\rm rot}/\sqrt{2(N-1)}$ ($\sigma/\sqrt{2(N-1)}$,
where $N$ is the total number of particles for a given radial bin.
We do not intend to separately estimate $V_{\rm rot}$ and $\sigma$ for
PNe originating from spiral's  stellar halos (thus metal-poor stars) 
and those from disks (metal-rich ones),
partly because observational data sets on metallicities of PNe
are not currently available. 
Although we find that there are no significant kinematical differences
between metal-poor and metal-rich PNe in some models, we will discuss 
possible metallicity dependences of PNe kinematics
in our forthcoming papers.

In order to investigate the smoothed 2D velocity fields,
we use the same smoothing method as that used for deriving the 2D
density fields of PNSs. 
The method of velocity smoothing is described as follows.
The particle with the location {\bf X},
and the velocity {\bf V}, and the mass of $M$ 
is considered to be composed of $N_{\rm s}$  particles (hereafter referred to
as ``smoothing particles'') that are located within $R$
(corresponding to smoothing length) from the particle (``parent particle'')
and have position vectors of ${\rm \bf x}_{i}$ ($i=1 \sim N_{\rm s}$) 
with respect to the parent,
velocity vectors of {\bf V} (i.e., the same as that of the parent), 
and masses of $M/N_{\rm s}$.  The spatial distribution
of these smoothing particles with respect to the parent particle ({\bf X})
follows the  Gaussian distributions. Therefore the probability of
a $i$-th smoothing particle with the position of  
${\rm \bf X} + {\rm \bf x}_{i}$ is proportional
to $\exp (-{r_i}^2/2R^2)$, where $r_i$ is the distance between the parent
particle and the smoothing one. If the number of smoothing particles for
a stellar particle is 100, the total number of ``particles''  used for
2500 bins is about $10^6$. Therefore, each bin includes 400 ``particles''
on average.

For each bin,  we average the velocities of ``particles'' with the positions
within the bin and derive the velocity at the location of the bin.
The adopted smoothing length is consistent with that used for observations
by PFF04a. Also we confirm that this method can reproduce
the ``spider shape'' (which is characteristic of rotational kinematics)
of the 2D velocity field of the initial disk. 
Thanks to this smoothing method, we can derive 2D density, velocity,
and dispersion fields of PNSs that can be tested against the corresponding
observations in a fully self-consistent manner.

This study is based totally on dissipationless simulations of galaxy
mergers so that it can not discuss structure and kinematics
of PNe formed from gas during/after galaxy merging. Accordingly,
the present results can be more reasonably compared with 
the observed 
E/S0s formed from ``dry'' (dissipationless) mergers that do not
create new PNe. In our future papers, we will discuss how the introduction
of PNe formation in our numerical simulations can change main
conclusions derived in the present study.

\section{Multiple mergers}

Each multiple merger model  contains equal-mass  spiral galaxies
with random orientations of intrinsic spin vectors which are 
uniformly distributed within  a sphere of size $6R_{\rm d}$.
The most important parameter in this multiple merger model
is the ratio of the initial kinematic energy ($T_{\rm kin}$) of the merger to 
that of initial potential ($W$).
By varying this ratio  ($t_{\rm v}$; defined as $|2T_{\rm kin}/W|$) from 0.25 to 0.75,
we investigate how  $t_{\rm v}$ controls the final PNe kinematics
of the merger remnants.
We mainly investigate
the results of two extreme cases: (1) where the initial kinetic energy
of a multiple merger is due entirely to the random motion of  
the five constituent galaxies (referred to as ``dispersion supported'')
and (2) where it is due entirely to
(rigid) rotational motion (``rotation supported'').
Our investigation of these two cases enables us to understand how
the initial rotation (dispersion) can control the final kinematical
properties of PNe in merger remnants.

Although we have derived the results of
three models with $t_{\rm v}$ = 0.25, 0.5, and 0.75
for each case (i.e., six models in total),
we describe the result of the ``rotation supported''
model with $t_{\rm v}$ = 0.5 (labeled as M10 in the Table 1). 
This is firstly because
this multiple merger model show some interesting differences in
PNe kinematics compared with pair merger models,
and secondary because this model shows  typical behaviors
in PNe kinematics among multiple merger models. 
The merger remnant of this  rotation supported model 
shows a flattened stellar (thus PNe) distribution if it is seen
from edge-on and have effective radius of 11.4 kpc for the stars
and 29.8 kpc for the dark matter halo. 
 
We briefly summarize the results  as follows.
The radial gradient in $\sigma$ is significantly shallower
in the multiple merger models than in pair merger models.
These  shallower $\sigma$ profiles in PNe kinematics
can be seen in most multiple  models and thus can be regarded
as one of characteristics of PNSs in multiple merger remnants.
The difference in $\sigma$
at each radial bin between PNe and dark matter halo
is slightly larger than that seen in pair merger models
 Both dark matter halo and PNe show
larger rotational velocities in the outer part of the halo 
and PNe can show larger $V_{\rm rot}$ than dark matter halo.

Thus PNSs in elliptical galaxies formed by multiple mergers
show similar kinematics to those in elliptical galaxies by pair mergers,
though the maximum values of $\sigma$ and $V_{\rm rot}$
can be different between the two cases owing to the larger masses
in multiple mergers.  Remarkable differences in PNe kinematics between
the two different merger models are (1) shallower slopes of radial  $\sigma$
profiles in multiple merger models and (2) larger $\sigma$ differences 
between dark matter halo and PNe for the multiple merger model.
The above  (1) suggests that there can be diversity in radial $\sigma$
profiles in elliptical galaxies, {\it if most of elliptical galaxies are
formed either from pair mergers or from multiple ones}.
The above (2) implies that radial $\sigma$ profiles of PNSs
alone can not allow us to correctly estimate of the total masses
of elliptical galaxies formed by multiple merging.
We have a plan to investigate this problem related to mass
estimation of galaxies by PNe kinematics in a more extensive 
manner in our forthcoming papers (Bekki \& Peng 2005)

\end{document}